\documentclass[journal]{IEEEtran}

\usepackage{amsmath,amsthm,amssymb}
\usepackage[compress]{cite}
\usepackage{tabto}
\usepackage{algorithm}
\usepackage[noend]{algpseudocode}
\usepackage{xspace}
\usepackage{paralist}
\usepackage{url}
\usepackage{graphicx}
\usepackage{tabularx,comment,makecell}
\usepackage{dsfont}
\usepackage[dvipsnames]{xcolor}
\usepackage{epigraph}
\usepackage{dblfloatfix}

\newcommand{\Fig}[1]{Fig.~\ref{fig:#1}}
\newcommand{\Prop}[1]{Property~\ref{prop:#1}}

\newcommand{\Sec}[1]{Sec.~\ref{sec:#1}}
\newcommand{\Tab}[1]{Tab.~\ref{tab:#1}}
\newcommand{\Eq}[1]{(\ref{eq:#1})}

\newcommand{\ee}{{\rm e}}
\newcommand{\jj}{{\rm j}}  
\newcommand{\dd}{{\rm\,d}} 

\newcommand{\Bc}{{\cal B}}
\newcommand{\Cc}{{\cal C}}
\newcommand{\Dc}{{\cal D}}
\newcommand{\Ec}{{\cal E}}
\newcommand{\Fc}{{\cal F}}

\newcommand{\Hc}{{\cal H}}
\newcommand{\Ic}{{\cal I}}

\newcommand{\Pc}{{\cal P}}

\newcommand{\Tc}{{\cal T}}

\newtheorem{property}{Property}
\DeclareMathOperator{\sgn}{sgn}

\begin{document}

\title{
Human-Centric Decision-Making\\in Cell-Less 6G Networks 
}

\author{
Emma~Chiaramello,~\IEEEmembership{Member,~IEEE,}
Carla~Fabiana~Chiasserini,~\IEEEmembership{Fellow,~IEEE,}
Francesco~Malandrino,~\IEEEmembership{Senior~Member,~IEEE,}
Alessandro~Nordio,~\IEEEmembership{Member,~IEEE,}
Marta~Parazzini,~\IEEEmembership{Member,~IEEE,}
Alvaro~Valcarce,~\IEEEmembership{Senior~Member,~IEEE}
\IEEEcompsocitemizethanks{\IEEEcompsocthanksitem
E.~Chiaramello, C.~F.~Chiasserini, F.~Malandrino, A.~Nordio and M.~Parazzini are with CNR-IEIIT and CNIT, Italy.
C.~F.~Chiasserini is with Politecnico di Torino, Italy and Chalmers University of Technology, Sweden. A.~Valcarce is with Nokia Bell Labs, France.
\IEEEcompsocthanksitem A preliminary version of this work has been presented in~\cite{noi-wowmom24}.
\IEEEcompsocthanksitem This work was supported by the SNS-JU-2022 project CENTRIC under the EU's Horizon Europe research and innovation programme, Grant Agreement No.\,101096379.
}
} 

\maketitle

\begin{abstract}
In next-generation networks, {\em cells} will be replaced by a collection of points-of-access (PoAs), with overlapping coverage areas and/or different technologies. Along with a promise for greater performance and flexibility, this creates further pressure on network management algorithms, which must make joint decisions on (i) PoA-to-user association and (ii) PoA management. We solve this challenging problem through an efficient and effective solution concept called Cluster-then-Match (CtM). Importantly, CtM makes {\em human-centric} decisions, where pure network performance is balanced against metrics like energy consumption and electromagnetic field exposure, which concern all humans in the network area -- including those who are not network users. Through our performance evaluation, which leverages detailed models for EMF exposure estimation and standard-specified signal propagation models, we show that CtM outperforms state-of-the-art network management schemes, including those utilizing machine learning, reducing energy consumption by over 80\%.
\end{abstract}


\section{Introduction}
\label{sec:intro}

\begin{figure}
\centering
\includegraphics[width=1\columnwidth]{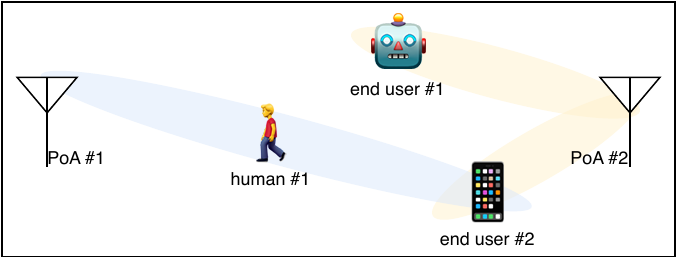}
\caption{
A simple indoor cell-less network. Two points-of-access (PoAs) serve their end users through one or more beams; one of the users can be served by two PoAs. A human is irradiated by the beam from PoA~\#1 to User~\#2;  humans will nonetheless incur EMF exposure whether or not they are network users.
    \label{fig:cellless}
}
\end{figure}

The recent trend in mobile communication networks is to make them  evolve away from the cellular paradigm. Indeed, traditional cells -- areas exclusively served by one base station, with overlap between cells being minimized -- are becoming increasingly rare. After the introduction of multi-layer coverage via small cells~\cite{nakamura2013trends} in 4G, and multi-RAT (radio access technology) networks in 5G~\cite{bangerter2014networks}, 6G will almost completely dispense with cells, and replace them with a collection of points-of-access (PoAs), thus becoming {\em cell less}. As exemplified in \Fig{cellless}, PoAs may use different technologies (e.g., 5G-NR and Wi-Fi) and overlapping coverage areas, and each end user (including user-terminals like smartphones, but also non-human devices like robots) can be served by multiple PoAs.

Another relevant feature of 5G-and-beyond networks is the increasing attention to sustainability, understood in a wide sense as the impact of networking on the environment; such impact can be evaluated through aspects like power consumption~\cite{ismail2014survey} and electromagnetic field (EMF) exposure~\cite{chiaramello2019stochastic}. This has generated a trend towards {\em human-centric} networking, a new paradigm where pure network performance (e.g., throughput) is balanced against sustainability metrics.
Compared to network performance metrics, human-centric ones tend to be more of a moving target: taking EMF exposure as an example, next-generation networks are moving towards higher frequencies, for which different metrics for EMF exposure assessment are being considered~\cite{zhekov2023study}. This means that our general goal is to endow human-centric network with a new capability -- {\em a priori} sustainability management --, rather than merely ensuring they comply with a static, given set of key performance and value indicator (KPI, KVI) targets.

As a result of the above, managing cell-less and human-centric networks will be substantially more complex than with current-generation ones. Reasons behind such complexity include (i) the number and diversity of the PoAs to manage; (ii) the fact that some technologies (especially higher-frequency ones) support beamforming~\cite{ahmed2018survey}, resulting in further decisions to make, and (iii) the fact that multiple PoAs may be available to serve a given user. Any decision made in cell-less and human-centric networks will impact, jointly and often in a counter-intuitive manner, both performance-related and human-centric metrics. The traditional (and popular, as detailed in \Sec{relwork}) approach of ignoring human-centric aspects is of course not an option; also, straightforward optimization approaches are ruled out by the complexity of the problem to solve, as per \Sec{anal}.

To address this issue, we propose an efficient solution strategy called {\em Cluster-then-Match} (CtM), which can jointly decide on the association between users and PoAs, and the management of the PoAs themselves, including beam steering and power management aspects at the PoAs. CtM improves over the state-of-the-art in three main ways, to wit:
\begin{enumerate}
    \item CtM jointly manages all PoAs, hence, it is able to prevent conflicts between different decisions;
    \item CtM considers all aspects of PoA management, from beam steering and beamwidth selection to power allocation;
    \item CtM considers  both performance-based and human-centric metrics, and the latter are evaluated for all humans in the network area, including those who are not network users.
\end{enumerate}
To ensure both the decision quality and the fact that they can be made swiftly enough, CtM performs two main steps: users are first {\em clustered} into groups, and then groups are {\em matched} with the PoAs. This allows us to exploit the flexibility of the cell-free networking paradigm, where the PoA (and beam) serving each end user can be chosen dynamically and there is no {\em a priori} association between users and PoAs.

Our contributions and structure of the paper can thus be summarized as follows:
\begin{itemize}
    \item We propose a concise and comprehensive model of cell-less networks, the decisions they require, and their effects on both performance-related and human-centric metrics, most notably, EMF exposure (\Sec{model});
    \item We formulate the resulting optimization problem, revealing how different decisions interact  (\Sec{problem}), and analyzing  its complexity, which rules out solving realistically-sized instances thereof (\Sec{anal});
    \item We present our CtM algorithm and highlight how it is able to ensure both the quality of network management decisions and the speed with which they are made (\Sec{algo});
    \item  We describe how to evaluate CtM's performance under standardized channel models and scenarios introduced in~\cite{ETSI_TR138901}, as well as detailed anatomic models for EMF exposure assessment (\Sec{scenario});
    \item We introduce the three benchmark we compare CtM against, including two based upon machine learning (ML), and present our performance evaluation results, showing that CtM outperforms the alternatives by over 80\% (\Sec{peva}).
\end{itemize}

\section{System Model}
\label{sec:model}

Our system model aims at representing the main elements of cell-less networks, their features, and the main decisions to make therein.

{\bf Model elements and parameters.}
Our system includes the PoAs, represented by elements~$p{\in}\Pc$, and the end users (either devices such as IoTs, or user terminals), represented by elements~$d\in\Dc$. Each PoA~$p{\in}\Pc$ is associated with one or more {\em beams}~$b{\in}\Bc$. Further, we consider a set of {\em humans}~$h{\in}\Hc$, representing people who might incur the EMF exposure resulting from the network. Humans in~$\Hc$ include both users of the network and simple passers-by, as depicted in \Fig{cellless}.

End users, humans, and PoAs, are associated with a position and height, described through parameters:
    $x_\text{H}(h)$, $y_\text{H}(h)$, and $z_\text{H}(h)$ for humans~$h\in\Hc$;
     $x_\text{D}(d)$, $y_\text{D}(d)$, and $z_\text{D}(d)$ for end users~$d\in\Dc$;
     $x_\text{P}(p)$, $y_\text{P}(p)$, and $z_\text{P}(p)$ for PoAs~$p\in\Pc$.
People with a user terminal (e.g., smartphone, tablet, laptop, wearable) are represented by {\em both} an end user in~$\Dc$ and a human in~$\Hc$, sharing the same location. IoT devices like robots, on the other hand, are only represented by an element in~$\Dc$.

Concerning PoAs~$p{\in}\Pc$, they operate at a  frequency~$f(p)$ (which is known) and feature a set of beams~$B(p){\subseteq}\Bc$, each of which must have a minimum width~$\omega^{\min}(p)$. Furthermore, we indicate with~$\pi(b){\in}\Pc$ the PoA originating beam~$b$. If a single PoA can operate at different frequencies, then two distinct elements in~$\Pc$ are created to represent it, sharing the same location.

{\bf Decisions and their effects.}
Our decisions concern the joint aspects of (i) end user-to-PoA assignment, and (ii) PoA and beam management. The former is addressed through binary variables~$y(b,d){\in}\{0,1\}$, expressing whether beam~$b{\in}\Bc$ (hence, PoA~$\pi(b){\in}\Pc$) serves device~$d{\in}\Dc$. Concerning the latter, we have to make four decisions, namely:
\begin{itemize}
    \item the transmission power~$P_\text{tx}(p)$ of PoAs~$p{\in}\Pc$;
    \item the azimuth angle~$\phi(b)$ of beam~$b{\in}\Bc$;
    \item its elevation angle~$\theta(b)$;
    \item its width~$\omega(b)$.
\end{itemize}

All the above decisions are then fed as input to the following three functions:
\begin{itemize}
    \item $\mathsf{Rate}(y,P_\text{tx},\phi,\theta,\omega,d)$, computing the data rate obtained by each end user~$d\in\Dc$;
    \item $\mathsf{Energy}(y,P_\text{tx},\phi,\theta,\omega)$, computing the total energy consumption;
    \item $\mathsf{Exposure}(y,P_\text{tx},\phi,\theta,\omega,h)$, estimating the EMF exposure levels incurred by each human~$h\in\Hc$, quantified through any of the metrics discussed in \Sec{exposure}.
\end{itemize}
These functions are then combined to express the objective we seek to optimize and the system constraints. From the viewpoint of our scheme, they are considered to be given and known.
It is important however  to remark how rate, energy, and exposure can be computed through different techniques -- and, in the case of exposure, even be quantified through different metrics. 

As an example, the received power can be determined through simple path-loss formulas or through much more complex channel models, accounting for clutter and mobility, such as the one we employ in \Sec{scenario}. Similarly, EMF exposure can be quantified through metrics that focus on the electromagnetic field at a given location (e.g., electric field strength or power density~\cite{colombi2015implications}), or through more complex metrics like the specific absorption rate (SAR), which account for the interaction between electromagnetic fields and biological tissues~\cite{hochwald2014incorporating}. Different scenarios and conditions call for different metrics, hence, there is no {\em right} way to compute the rate, energy, and exposure.

Accordingly, our system model and problem formulation -- as well as the CtM solution strategy 
-- can accommodate {\em any} technique to do so, from the simplest to the most realistic ones. This, coupled with the modular architecture of our implementation (see \Sec{archi}), greatly improves the flexibility of the proposed CtM solution, as well as its applicability to many real-world scenarios and conditions.

\section{Optimization Formulation}
\label{sec:problem}

This section first presents the formulation of the problem we address, and then proves its NP-hardness.

\subsection{Objective and requirements.\label{sec:problem-f}}
In general, we  aim to optimize one of the metrics introduced above, i.e., one of the functions~$\textsf{Rate}$, $\textsf{Energy}$ or~$\textsf{Exposure}$, while using the other two as constraints. The traditional approach would be to maximize the performance (quantified, typically, through the sum-rate~\cite{sediq2013optimal}), subject to constraints about energy and EMF exposure. In contrast, following the human-centric paradigm, we seek to minimize the energy, subject to the fact that (i) each end user~$d{\in}\Dc$ is guaranteed the required rate, and (ii) exposure limits are honored for all humans in~$\Hc$.

More formally, we have:
\begin{align}
\label{eq:obj} \min_{y,P_\text{tx},\phi,\theta,\omega} & \textsf{Energy}(y,P_\text{tx},\phi,\theta,\omega) & \\
\label{eq:rat}  & \textsf{Rate}(y,P_\text{tx},\phi,\theta,\omega,d) \geq \textsf{Rate}^{\min}(d), \forall d\\
\label{eq:exp}  & \textsf{Exposure}(y,P_\text{tx},\phi,\theta,\omega,h) \leq \textsf{Exposure}^{\max},\forall h.
\end{align}
The problem above reflects the complexity of our scenario, as discussed earlier:  the objective \Eq{obj} concerns a global variable, the rate requirements are to be guaranteed for each single end user, while meeting the guidelines about EMF exposure has to be ensured for all humans.

Also notice how minimum rates may be different for all end users, reflecting the fact that performance
requirements vary wildly across applications and services. With regard to the EMF exposure, the international guidelines have been defined by international scientific bodies, e.g., the International Commission on Non-Ionizing Radiation Protection (ICNIRP), to protect people from all substantiated harmful effects of EMF exposure. The estimation of the levels of EMF exposure (i.e.,  the~$\mathsf{Exposure}$ function in this study) is performed here in terms of the rate of energy absorbed by the biological tissues; thus, its values will be different for humans with different morphologies, consistently with the human-centric approach of our study. We remark that the levels of EMF exposure is not used as optimization objective, as the goal is not to obtain the lowest level possible, but to be compliant with the international exposure guidelines.

\subsection{Problem analysis}
\label{sec:anal}
We can prove the following property.
\begin{property}
\label{prop:hard}
The problem of minimizing \Eq{obj} subject to \Eq{rat} and \Eq{exp} is NP-hard.
\end{property}
\begin{IEEEproof}
We show how  any instance of the generalized assignment problem (GAP), which is known to be NP-hard~\cite{cattrysse1992survey}, can be  transformed into an instance of our problem. 
In GAP,  a set of~$\Ic{=}\{i\}$ of {\em items}  must be  assigned  to  containers (bins)~$c{\in}\Cc$. Items have a {\em value}~$v_i$ and a weight~$w_i$; containers can support a maximum weight of~$W_c$. 
Any GAP instance can be transformed in a heavily simplified instance of our problem, where:
    
    $\bullet$ PoAs in~$\Pc$ correspond to items in~$\Ic$;
    
   $\bullet$ end users in~$\Dc$ correspond to containers in~$\Cc$;
   
   $\bullet$ 
   $|\Bc|$ is such that all PoAs can serve all end users;

   $\bullet$ the rate experienced by end user~$d$ when served by PoA~$p$ is equal to~$R(p,d)$, regardless of all other decisions, and such a quantity is equal to the weight of the item corresponding to the PoA;
   
    $\bullet$ the function $\mathsf{Rate}$ is such that end users served by multiple PoAs simply enjoy the sum of the individual rates~$R(p,d)$, i.e., there is no scheduling and no interference;
        
     $\bullet$ no EMF exposure is incurred (or, alternatively, there are no humans in~$\Hc$).

\noindent
Notice that these are highly simplified  assumptions, i.e., we are transforming a full instance of GAP into a very simple (hence, intuitively, relatively easy to solve) instance of ours.

We start from a condition where all PoAs serve all end users, and assign to each end user a minimum rate of
$\textsf{Rate}^{\min}(d)=\sum_{p\in\Pc}\mathsf{Rate}(p,d)-W_c,$
where~$c$ is the container associated with end user~$d$. Similarly to the rate, serving end user $d$ with PoA~$p$ incurs an energy cost of~$E(c,d)$, and the total energy (i.e., the value of objective \Eq{obj}) is simply given by~$\sum_{c,d}E(c,d)$.
Also, initially, all PoAs serve all end users with one beam. Then, assigning item~$i$ to container~$c$ (in the GAP) means {\em stopping} serving the end user~$d$ associated through~$c$ with the PoA associated with~$i$. Thus, every time we assign an item, we decrease the energy cost in our problem (which corresponds to increasing the value in the GAP). At the same time, the rate obtained by the end user  decreases (i.e., the weight in the container increases) and gets closer to the minimum rate (i.e., the maximum  container weight). 
Finally, the optimal solution of our problem 
is also the optimal solution for the GAP.  
Through the above procedure, we can reduce any instance of the GAP problem to an instance of ours. The reduction requires polynomial (linear) time, hence, the thesis is proved.
\end{IEEEproof}
There are two aspects of the proof of \Prop{hard} that are worth highlighting. First, we convert an instance of the GAP problem to a {\em very simple} instance of our own; intuitively, this suggests that our problem might be even more complex than GAP. Furthermore, the proof makes heavily simplified assumptions on the form of the functions~$\mathsf{Rate}$ and $\mathsf{Energy}$, while altogether ignoring~$\mathsf{Exposure}$. As exemplified by the channel model presented in \Sec{scenario}, these functions can be very complex to compute, non-linear, and non-convex; indeed, they may even lack a closed-form expression. All such elements add further complexity to our problem.

As a consequence of \Prop{hard} and the remarks above, it is impractical -- and, often, impossible -- to solve our problem using conventional numerical solvers. This motivates us to propose an efficient and effective heuristic strategy called cluster-then-match (CtM), as described next.

\section{The CtM Strategy}
\label{sec:algo}

As discussed in \Sec{anal}, it is prohibitively complex to solve our problem to optimality, mostly due  to the size of its {\em solution space}. Accordingly, our CtM heuristic strategy is predicated on (i) sequentially considering the main elements of our system model (end users, PoAs, humans), and (ii) at each step, restricting our attention to the most promising possible decisions. In doing so, CtM can explore a {\em subset} of the original solution space that, nonetheless, contains most of (often, all) the highest-quality solutions.

\begin{figure}
\centering
\includegraphics[width=1\columnwidth]{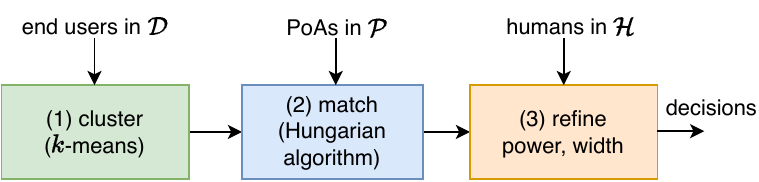}
\caption{
The three main steps of the CtM strategy: {\em clustering} of the end users, cluster-to-PoA {\em assignment}, {\em optimizing} beam width and transmission power levels.
\label{fig:ctm}
} 
\end{figure}

Specifically, as depicted in \Fig{ctm}, the CtM strategy consists of three main steps:
\begin{enumerate}
    \item clustering the end users in~$\Dc$, so as to  make subsequent decisions on a per-cluster, rather than per-end user, basis;
    \item assigning clusters to PoAs in~$\Pc$, thus determining beam steering;
    \item optimizing beam width and transmission power levels.
\end{enumerate}
In the following, we describe each of the steps separately.

{\bf Step~1: Clustering.}
A major reason for the complexity of the original problem formulated in \Sec{problem} is the many-to-many relationship between end users in~$\Dc$ and PoAs in~$\Pc$. On the one hand, we need to take this relationship into account  to exploit the potential of cell-less networks; on the other hand, considering {\em all} possible end user-to-PoA associations is unnecessarily complex. Our intuition is to leverage the beams in~$\Bc{=}\bigcup_{p\in\Pc}B(p)$, and make the key observations that:

    $\bullet$ since there will be more end users than beams, each beam will serve multiple end users, and
    
    $\bullet$ to keep beams as narrow as possible, it is desirable that end users served by the same beam are close to each other.
The latter is motivated by the fact that wider beams result in more interference, as well as higher energy consumption. 

Following the observations above, in Step~1 of the CtM strategy we {\em cluster} the end users in~$\Dc$ into as many clusters as the number of possible  beams in~$\Bc$, exploiting their position information (i.e., the~$x_\text{P}$ and~$y_\text{P}$ parameters) to make clusters as small -- in terms of area occupied by their end users -- as possible. Although  any clustering algorithm can be used in this step, for concreteness we adopt the  $k$-means algorithm~\cite{hartigan1979algorithm}. $k$-means indeed yields very good results in scenarios like ours, also thanks to the fact that (unlike hierarchical clustering and later approaches such as DBSCAN~\cite{schubert2017dbscan}) it takes as an input the target number~$k$ of clusters, which corresponds to the number of beams in our case. End users in the same cluster can be served either via pairing in multi-user, multiple-input, multiple-output MIMO (MU-MIMO) systems~\cite{sur2016practical}, or via time-division multiplexing (TDM) if MU-MIMO is not used in the current scenario.

{\bf Step~2: Matching.}
We have now as many clusters of end users as the number of beams in~$\Bc$; our next task is to {\em match} the beams and the clusters they serve. This is a {\em bipartite matching} problem; importantly, thanks to the clustering performed in Step~1, the matching  to perform is one-to-one instead of many-to-many. To efficiently solve to optimality our one-to-one bipartite matching problem, we apply the Hungarian algorithm~\cite{jonker1986improving}. The input to the algorithm consists of a square matrix~$\mathbf{C}{=}\{c_{kb}\}$, where element~$c_{kb}$ is the {\em cost} of serving the $k$-th cluster of devices through the $b$-th beam; the output is the assignment minimizing the sum of the costs.

As our cost metric, we simply consider the {\em distance} from the PoA originating the beam to the centroid of the cluster of devices to serve. Making decisions based upon the distance has a twofold advantage:
\begin{itemize}
    \item shorter distances between PoAs and devices make it possible to use lower power levels and narrower beams;
    \item using distances (as opposed to, e.g., the achievable rate) does not require any knowledge of the specific implementation of the~$\mathsf{Energy}$, $\mathsf{Rate}$, or~$\mathsf{Exposure}$ functions.
\end{itemize}
The latter item also implies that Step~2 of CtM is easy and quick to perform even in scenarios and cases where evaluating the rate, energy, or exposure is costly and/or time consuming (e.g., if those quantities are evaluated through simulations).

{\bf Step~3: Optimizing beam width and power.}
In this step, we set the width of each beam and the transmission power of each PoA, keeping into account both the objective \Eq{obj} and the constraints \Eq{rat} and \Eq{exp}. Specifically, we set the width of beams to the minimum value necessary to serve the end users assigned to them. Additionally,  we reduce their power as much as possible to decrease the energy consumption and the EMF exposure, while ensuring the required performance.

Concerning beam widths, let~$D(b){\subseteq}\Dc$ be the set of end users belonging to the cluster assigned to beam~$b$. Then the angle~$\alpha(b,d)$ from each end user~$d{\in} D(b)$ is given by:
\begin{equation}
\nonumber
\alpha(b,d) = \sgn(\Delta_y) \arccos \frac{\Delta_x}{\sqrt{\Delta_x^2+\Delta_y^2}},
\end{equation}
where~$\Delta_y {=}y_D(d){-}y_P(\pi(b))$, $\Delta_x {=}x_D(d) {-}x_P(\pi(b))$
and the width of the beam is set to angular width of the cluster, as observed from the PoA generating beam $b$, i.e.,
\begin{equation}
\nonumber
\omega(b)\gets \max\left\{\omega^{\min}(\pi(b)),\pi-\left|\alpha(b)^{\max}-\alpha(b)^{\min}-\pi\right|\right\},
\end{equation}
where $\alpha(b)^{\max}{=}\displaystyle\max_{d\in D(b)}\alpha(b,d)$ and $\alpha(b)^{\min}{=}\displaystyle\min_{d\in D(b)}\alpha(b,d)$.
Notice how the above expression also accounts for the minimum beam width~$\omega^{\min}$, reflecting the fact that there are technological limits -- specific to each PoA -- preventing beams from being as narrow as one might desire.

The last part of Step~3 concerns power levels. Our key observations are as follows:
\begin{itemize}
    \item reducing power might endanger feasibility (by resulting in a violation of constraint \Eq{rat}), but will never affect adversely energy consumption \Eq{obj} or EMF exposure \Eq{exp};
    \item reducing the power of a PoA will not impact feasibility for end users served by another PoA.
\end{itemize}
Accordingly, we process one PoA at a time,  as follows:
\begin{enumerate}
    \item[(a)] set an amount of power~$\delta$;
    \item[(b)] consider one of the PoAs~$p{\in}\Pc$;
    \item[(c)] reduce the power~$P_\text{tx}(p)$ by~$\delta$;
    \item[(d)] repeat Step~(c) so long as that results in a feasible solution;
    \item[(e)] go back to Step~(b) and consider a different PoA.
\end{enumerate}
The above procedure can be repeated, in a manner similar to Newton's bisection algorithm~\cite{kearfott1987some}, setting a smaller~$\delta$ and starting afresh from Step~(a). By doing so, we can get even more fine-grained (hence, higher quality) decisions, at the cost of a longer running time.

\subsection{Algorithm analysis}
\label{sec:aanal}

 We now formally prove that CtM makes high-quality decisions {\em swiftly}, by having a low computational complexity regardless the size of the problem to solve.
\begin{property}
The CtM strategy has polynomial worst-case computational complexity.
\label{prop:easy}
\end{property}
\begin{IEEEproof}
We prove the property by separately considering the three steps depicted in \Fig{ctm}. 
Concerning Step~1, virtually all popular clustering techniques have polynomial complexity; taking $k$-means in particular, recent implementations can even achieve linear complexity~\cite{pakhira2014linear}. 
The matching in Step~2 can be performed optimally through multiple approaches; considering the Hungarian algorithm, its complexity is cubic in the number of items to match (in our case, beams), as shown in~\cite{edmonds1972theoretical}. 
In Step~3, width decisions are made only once per beam, hence, the added complexity is linear in the number of beams. Power levels are instead reduced at most~$|\Pc|\left\lceil\frac{P_\text{tx}^{\max}}{\delta}\right\rceil$ times, hence, the added complexity is linear in the number of PoAs.

As each of the above steps has polynomial time complexity, and we perform them sequentially, it follows that the total complexity of the full CtM procedure is polynomial as well.
\end{IEEEproof} 
The concrete meaning of \Prop{easy} is that the computational complexity, hence, the time required by CtM to make its decisions, remains manageable as the size of the problem instance (e.g., the number of end users and PoAs) grows. This, in turn, implies that CtM decisions are made in time to be useful (e.g., to be acted upon) even in large and/or complex, real-world scenarios.

\begin{figure*}[t]
\centering
\includegraphics[width=0.58\textwidth]{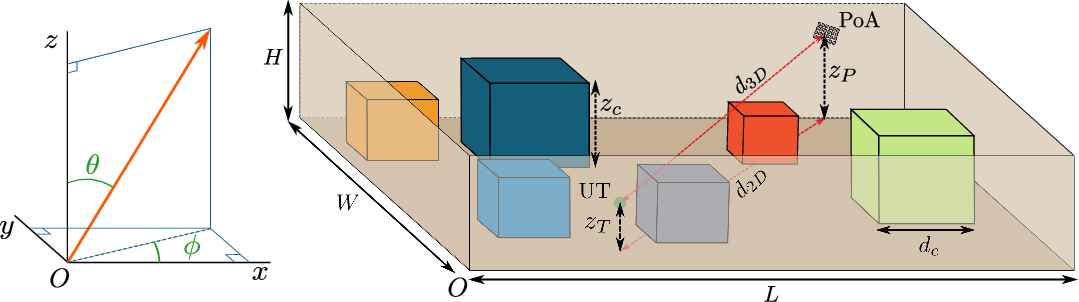}
\hspace{4mm}
\includegraphics[width=0.22\textwidth]{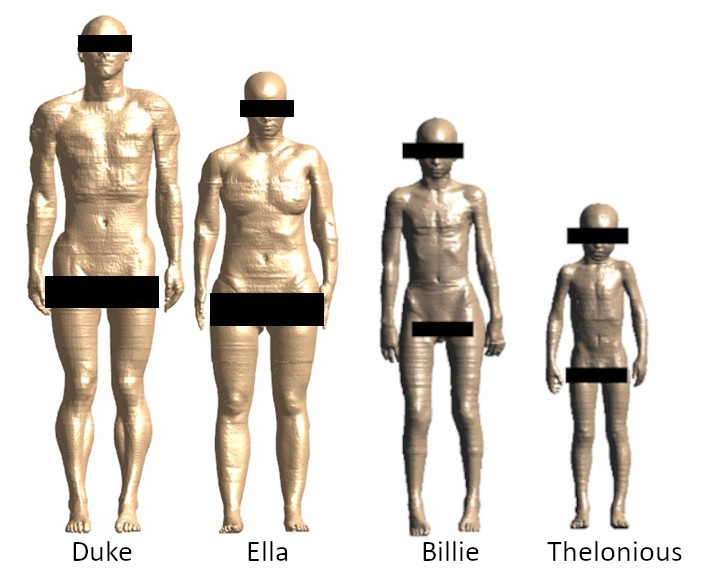}
\caption{Global coordinate system (GCS) (left),  hall model for the InF-DH indoor scenario with colored objects representing  clutters (center), and the detailed anatomical models used for electromagnetic field exposure assessment (right).}
\label{fig:hall}
\end{figure*}

\section{Reference Scenario}
\label{sec:scenario}

We consider two reference scenarios, to wit, an indoor factory (InF-DH) and and an outdoor urban canyon (UMI-SC). Each scenario includes a set of PoAs, $\Pc$, of end users, $\Dc$, and of humans $\Hc$. Furthermore, indoor scenarios may include walls, ceiling, and clutter, i.e., metallic machinery or other irregular objects, as exemplified in \Fig{hall}(center).

The signal propagation in these environments depends upon several factors, including the frequency used for the communication, the orientation of antennas and beams, and the position of PoAs and users. Indoor scenarios also account for the hall size and the clutter density. The ETSI technical report~\cite{ETSI_TR138901} contains a comprehensive set of propagation models, tailored for different scenarios and environments.

The environment is associated with a spherical or Cartesian global coordinate
system (GCS), as depicted in \Fig{hall}(left) for the indoor scenario. In the GCS, the azimuth angle,
$\phi\in[-180^\circ,180^\circ]$, is measured counterclockwise from the
$x$ axis, whereas the zenith angle, $\theta\in[0^\circ,180^\circ]$, is
measured from the $z$ axis. Thus, directions on the horizon have
$\theta{=}90^\circ$.  To characterize the propagation channel, we first
discuss the antenna model and then we  detail  the channel
impulse response.

\subsection{Antenna model and beamforming}
The ETSI specifications~\cite{ETSI_TR138901} assume that PoAs and receivers are
equipped with rectangular arrays (antenna panels) of $M{\times} N$  elements  organized in a regular grid. 
To each panel is associated a local coordinate
system (LCS), 
with the $x$ axis
being along the panel broadside direction. The mechanical orientation of
the panel thus completely defines the relation between its LCS and the GCS. 
To simplify  the model description, we here
assume that all panels have tilt angle equal to $90^\circ$ and slant
angle $0^\circ$, whereas the azimuth angle, $\alpha$, can be arbitrarily
chosen. 
The radiation power pattern of a panel element can be described in
the LCS by the function $A_{\rm dB}(\theta,\phi)$\footnote{A variable name with value expressed in dB 
shows the subscript ``dB''.},
given by 
 $ A_{\rm dB}(\theta,\phi) {=}  0$
the ideal isotropic model, and by 
$ A_{\rm dB}(\theta,\phi) {=} 8-\min\{A_{1,{\rm dB}}(\theta)+A_{2,{\rm dB}}(\phi),30\} $ 
with 
$  A_{1,{\rm dB}}(\theta) {=} \min\{12(\frac{\theta-90^\circ}{65})^2,30\} $
and
$A_{2,{\rm dB}}(\phi)   {=} \min\{12(\frac{\phi}{65})^2,30\}$, 
according to the 3GPP model with 8\,dBi gain.

An antenna panel can 
generate a beam $b$ toward a desired direction, characterized by the angles
$(\tilde{\theta},\tilde{\phi})$, measured w.r.t. the LCS. 
%
Assuming that all antenna elements are vertically polarized, the
electromagnetic field sent by the panel along the  direction
$(\theta,\phi)$ (w.r.t. the LCS) is given by
\begin{eqnarray}
\hspace{-0.2cm} F(\theta,\phi,\tilde{\theta},\tilde{\phi},M,N) 
 & {=} &
\widetilde{F}(\theta,\phi)\frac{{\rm sinc}(M g_1)}{{\rm sinc}(g_1)}\frac{{\rm sinc}(N g_2)}{{\rm sinc}(g_2)} \cdot \nonumber \\
&&  \sqrt{NM}\ee^{\jj\pi[(M-1)g_1+(N-1)g_2]} \label{eq:field} 
  \nonumber
\end{eqnarray}
where $\widetilde{F}(\theta,\phi){=}10^{A_{\rm dB}(\theta,\phi)/20}$ is the field generated by an array element,  
and
 $ g_1 
 {=} d_{\rm v}(\cos \theta - \cos \tilde{\theta})$ 
and $  g_2 
{=} d_{\rm h}(\sin \phi \sin \theta - \sin \tilde{\phi} \sin \tilde{\theta})$.
Note  that such an antenna model can  describe both the
field associated with antennas used for communication as well as a
human body.  In the latter case, a portion of a body with surface
$\lambda^2/4\pi$\,m$^2$ can be associated with a single isotropic
antenna element with gain given by $A_{\rm dB}(\theta,\phi){=}0$ and effective area $\lambda^2/4\pi$, where $\lambda$ is the signal
wavelength.

\subsection{Channel impulse response\label{sec:channel}}
We now focus on a single wireless link connecting a generic PoA $p{\in}
\Pc$ to a generic target $t{\in} \Tc$ where $\Tc{=}\Dc {\cup} \Hc$ is the
set comprising end users and humans. For simplicity, we omit the dependence on  PoA $p$ and target
$t$. The distance between the PoA and the target is
denoted by $d_{\rm 3D}$, and its projection on the ground is
referred to as $d_{\rm 2D}$ (see \Fig{hall}). The InF-DH and UMI-SC model provide expressions for the channel impulse response,
$\tilde{h}(\tau)$, which accounts for both line of sight (LoS) and
non-LoS (nLoS) propagation. The probability of the PoA and the target
being in LoS depends on several factors, including the distance $d_{\rm 2D}$, on the PoA and
target height and (for the InF-DH scenario) on the clutter density
\cite{ETSI_TR138901}.
The channel impulse response has, in general, the following expression:
$ \tilde{h}(\tau) = 10^{(\gamma_{\rm dB}+P_{\rm tx,
      dB})/20+\zeta}h(\tau) $ where $P_{\rm tx,dB}$ is the power
emitted by the PoA $p$ (in dBm), $\gamma_{\rm dB}$
is the pathloss, and $\zeta$ is a random variable modeling shadow
fading. 
The function $h(\tau)$ describes the wireless link as a set of rays
grouped in a (scenario-specific) number $N_c$ of clusters. Cluster $j$ is described by the
normalized power $P_j$ and collects the contribution of $N_r{=}20$ rays,
which are characterized by a common delay $\tau_j$.  Ray $\ell$ of
cluster $j$ has azimuth and zenith angles of departure (AoD)
$(\theta_{P,j,\ell},\phi_{P,j,\ell})$, measured w.r.t. to the GCS and,
similarly, azimuth and zenith angles of arrival (AoA)
$(\theta_{T,j,\ell},\phi_{T,j,\ell})$, when observed from the
target. If the PoA and the target are in LoS, an additional
path, with LoS AoD $(\theta_{P,0},\phi_{P,0})$ and LoS AoA
$(\theta_{T,0},\phi_{T,0})$ (in the GCS), has to be considered.  Thus, in the delay domain, we have: 
\begin{equation}
h(\tau) = \left\{
  \begin{array}{ll}
     \sum_{j=1}^{N_c}\sqrt{\frac{P_j}{N_r}}\sum_{\ell=1}^{N_r}
F_{T,j,\ell}F_{P,j,\ell}\ee^{\jj\varphi_{j,\ell}}\delta(\tau{-}\tau_j) & \\ \qquad\qquad\mbox{ if nLoS propagation
       apply} \\
     \sqrt{\frac{1}{1+K}}h^{\rm nLoS}(\tau)+\sqrt{\frac{K}{1+K}}h^{\rm
       LoS}\delta(\tau-\tau_1) & \\ \qquad\qquad \mbox{ if LoS
       propagation apply}
  \end{array}
\right.\label{eq:h} \end{equation}
where $\delta(\tau)$ is the Dirac delta function,  $K$ is the Rician fading coefficient, 
$\varphi_{j,\ell}$ is a random phase and
$F_{T,j,\ell} {\triangleq} F(\theta_{T,j,\ell},\phi_{T,j,\ell}{-}\alpha_T,\tilde{\theta}_T,\tilde{\phi}_T,M_T,N_T) $
 and $F_{P,j,\ell} {\triangleq} F(\theta_{P,j,\ell},\phi_{P,j,\ell}{-}\alpha_P,\tilde{\theta}_P,\tilde{\phi}_P,M_P,N_P)$ 
are the PoA and target fields computed according
to~\eqref{eq:field}.
Specifically, for computing $F_{T,j,\ell}$, we substituted
in~\eqref{eq:field} the antenna parameters of the target (e.g., $M_T$, $N_T$,
and beam direction $\tilde{\phi}_T, \tilde{\theta}_T$ while,  for computing~$F_{P,j,\ell}$,  
we employed those of the PoA, namely, $M_P$, $N_P$, $\tilde{\phi}_P$ and $\tilde{\theta}_P$.
Moreover, in such expressions, the azimuth angles $\alpha_P$ and
$\alpha_T$ represent the azimuth mechanical orientation of the PoA and target
antennas, respectively. Finally,
$  h^{\rm LoS}= F_{T,0}F_{P,0}\ee^{-\jj\frac{2\pi}{\lambda}d_{\rm 3D}}$ 
with  
$F_{T,0} {\triangleq} F(\theta_{T,0},\phi_{T,0}{-}\alpha_T,\tilde{\phi}_T,\tilde{\theta}_T,M_T,N_T)$ 
and 
$F_{P,0} {\triangleq} F(\theta_{P,0},\phi_{P,0}{-}\alpha_P,\tilde{\phi}_P,\tilde{\theta}_P,M_P,N_P)$.  
Note that, according to~\cite{ETSI_TR138901}, the values of the Rician
fading coefficient $K$, of the AOA and AOD of each ray, of the cluster
powers $P_j$ and of the delays $\tau_j$ take random values and are obtained through a long
and complex procedure whose description would require a large amount of space. The interested reader is referred to~\cite{ETSI_TR138901}
for details. 
If in the scenario there are several PoAs and targets, the wireless
links connecting any pair of them are to be considered mutually
independent.

 \subsection{SINR, rate, and power density at the target}
  Given  the set of simultaneously transmitting PoAs ($\Pc$) and 
    the set of targets ($\Tc$), we recall that with  each PoA $p$ is
  associated a position $[x_P(p),y_P(p),z_P(p)]$, a carrier frequency
  $f(p)$, a panel with $M_P(p)\times N_P(p)$ elements, and a mechanical
  orientation angle $\alpha_P(p)$, a transmit power $P_{\rm
    tx,dB}(p)$, and a beam $b$, whose direction, in the panel LCS, is
  specified by the angles $\tilde{\phi}_P(p)$ and
  $\tilde{\theta}_P(p)$. Similarly, the target has position
  $[x_T(t),y_T(t),z_T(t)]$. For simplicity, we assume that the target
  has an antenna panel with $M_T(t){=}N_T(t){=}1$ element and its power
  pattern is isotropic. 
  Thus, the target 
  mechanical orientation angle becomes meaningless as well as the
  notion of beam, i.e., the variables $\alpha_T(t)$,
  $\tilde{\phi}_T(t)$, and $\tilde{\theta}_T(t)$ can be arbitrarily
  chosen since they have no effect. 
  Then, for every $p$ and $t$,
  we define the impulse response of the channel connecting them 
  as $\tilde{h}_{p,t}(\tau)$, computed as specified in Sec.~\ref{sec:channel}, with power 
  $\Ec_{p,t} {=} \int_{-\infty}^{+\infty}|\tilde{h}_{p,t}(\tau)|^2\dd \tau$.
  Let $\Fc$ be the set of frequencies used by the PoAs and  $\Pc(f)\subseteq \Pc$ the subset
  of PoAs transmitting at frequency $f {\in} \Fc$, where $\cup_{f \in \Fc}
  \Pc(f) {=} \Pc$ and $\Pc(f) {\cap} \Pc(f') {=} \emptyset$ $\forall\, f,f'{\in}\Fc$.

  Thus, when $p$ transmits and the target $t$ is an end user $d$, the instantaneous measured SINR is
  \[ {\rm SINR}_{p,d} = \frac{\Ec_{p,t}}{N_0 W_p+ \int_{-\infty}^{+\infty}\left|\sum_{p'\in \Pc(f(p)),p'\neq p}\tilde{h}_{p,d}(\tau)\right|^2\dd\tau}\,,\]
  where $W_p$ is the signal bandwidth used by $p$ and $N_0{=}-174$\,dBm/Hz is the
  thermal noise power spectral density. The corresponding achievable
  rate is then given by $R_{p,d}{=}W_p\log_2(1+ {\rm SINR}_{p,d})$. 
  Instead, when the target $t$ is a human $h$, the power received from all transmitting PoAs at frequency $f$ is: 
  $P_{{\rm rx},h,f} {=} \int_{-\infty}^{+\infty}\left|\sum_{p\in \Pc(f)}\tilde{h}_{p,h}(\tau)\right|^2\dd \tau$
  and the corresponding power density over the human body at $h$ is obtained by dividing the received power in a certain
point by the effective area $\lambda^2/(4\pi)$, i.e., $S_{h,f} {=}  4\pi f^2/c^2 P_{{\rm rx},t,f}$\,W/m$^2$.

\subsection{Exposure model}
\label{sec:exposure}
Given the power density over the human body $S_{h,f}$, 
the EMF exposure is assessed by estimating the specific energy
absorption rate (SAR) over the whole body, SAR$_{\rm wb}$, which is the
power of the EMF absorbed over the entire body mass. 
We first derive the incident electric field $E_{\rm inc}{=} \sqrt{S_{h,f} Z_0 }$\,V/m 
where $Z_0{=}377\,\Omega$ is the free-space impedance. Considering a
human with body mass index ${\rm BMI}_h$\,kg/m$^2$, the SAR$_{\rm wb}$ can be
estimated as
\begin{equation}\label{eq:2EMF}
{\rm SAR}_{\rm wb} {=} \left(\frac{E_{\rm inc}}{E_{\rm ref}}\right)^2\cdot \frac{{\rm BMI}_h}{{\rm BMI}_{\rm ref}}{\rm SAR}_{\rm ref}
\end{equation}
where SAR$_{\rm ref}$, measured in W/kg is the whole-body SAR induced
by a reference incident field $E_{\rm ref}$ in a reference human body
of mass index BMI$_{\rm ref}$. Eq.~\eqref{eq:2EMF} assumes
far-field conditions, 
which is a reasonable assumption for the exposure
scenarios and the frequency considered in this study. 

The whole-body specific absorption rate (SAR$_{\rm ref}$) of the
reference human body 
is defined
as~\cite{international2020guidelines}:
${\rm SAR}_{\rm ref} {=} \frac{P_{\rm wb,ref}}{M_{\rm ref}} {=} \frac{1}{M_{\rm ref}} \int_{\rm wb} \sigma (r) E^2_{\rm RMS}(r) \,dV$ 
where $P_{\rm wb,ref}$ (in W) and $M_{\rm ref}$ (in
kg) are, respectively, the whole-body absorbed power and the mass of
the reference human body. As for the reference human body: 
(i) $\sigma$ (in S/m) is its electrical conductivity;
(ii) $E_{\rm RMS}$ (in V/m) is the root mean square value of the induced electric field, 
  and (iii) $V$ is its volume.
The values of SAR$_{\rm ref}$ considered in this study were calculated by~\cite{liorni2020} through
electromagnetic computational techniques in anatomical human models and determined in far-field conditions, when the reference field
$E_{\rm ref}$ was set equal to 2.45\,V/m.
To account for the anatomical
variability of human bodies,  we estimated the SAR$_{\rm wb}$ for
four computational whole-body anatomical human models, two adults and two children: Ella (female, 26
years), Duke (male, 34 years), Thelonious (male, 6 years) and Billie (female, 11 years) (see Fig.\,\ref{fig:hall}(right)).
These models belong to the ``Virtual Population'', a collection of anatomical
models obtained from high-resolution Magnetic Resonance Imaging (MRI)
data widely used for electromagnetic field exposure
assessment~\cite{christ_virtual_2009}. Each model
resembled the anatomical characteristics of a real individual and included up to more than 80 different tissues across the whole body.

The SAR$_{\rm ref}$ values provided by~\cite{liorni2020} were
calculated at a set of frequencies slightly different from those
considered in this study, i.e., 3 and 5~GHz. Since the dielectric properties of human tissues vary with frequency (thus affecting the absorption of the power by the human tissues),
we needed to identify which value of SAR$_{\rm ref}$ could be considered a reliable
approximations of those obtained 
at precisely 3 and 5\,GHz. To do so, we computed the ratio between the dielectric properties
across all the tissues, at 3 and 5\,GHz and those at the nearest
frequencies reported in~\cite{liorni2020}, i.e., 2.45, 3.5 and 5.2\,GHz. All   obtained values were well below 2, the
threshold for which, variations in
dielectric properties do not substantially influence the whole-body
SAR~\cite{Gajsek2001}. For 5\,GHz, the dielectric properties of human
tissues~\cite{Gabriel_1996A, Gabriel_1996B} were very similar to those
reported at 5.2\,GHz. 
As to the 3\,GHz frequency, the
dielectric properties were compared with those reported for both 2.45
and 3.5\,GHz, to identify which approximation could be the best. Although for both frequencies the ratios obtained as average
across all the tissues were well below 2, the best approximation was
found to be the 2.45\,GHz.
Evaluating SAR values
allows  assessing the compliance of the dose of EMF absorbed with the
basic restrictions limits recommended by the guidelines set by
ICNIRP~\cite{international2020guidelines}. For the whole body SAR
the basic restrictions for general public are equal to 0.08\,W/kg for
both the considered frequencies~\cite{international2020guidelines}.

\subsection{Reference topologies}
\label{sec:scenarios}
We consider two scenarios, both drawn from ETSI technical report~\cite{ETSI_TR138901}:
\begin{itemize}
    \item An indoor factory scenario (Indoor Factory with Dense clutter and High base station height in~\cite{ETSI_TR138901}, or InF-DH), considering a $20\times 80$~m$^2$ factory hall where $\rho_c=40\%$~of the floor space is occupied by clutter, whose typical height is~$2$~m;
    \item An urban canyon scenario (UMI-SC in~\cite{ETSI_TR138901}), describing a $40\times 800$~m$^2$ road stretch.
\end{itemize}

For EMF exposure estimation, we consider the four human anatomical models described above, consisting of two adults, one adolescent and one child.
Using models of people of different age and body structure allows us to enhance the human-centric aspect of our work, by more accurately estimating the effect of networks over the entire population. Notice that we only consider models of adults (Ella and Duke) for the indoor factory scenarios, as the youngest individuals are unlikely to be found there; on the other hand, we use all models for the urban canyon scenario. Also notice how, as per~\cite{ETSI_TR138901}, the urban canyon requires a minimum of 10~meters distance between users and the PoAs serving them; humans, on the other hand, can be placed anywhere in the topology.

\section{Performance Evaluation}
\label{sec:peva}

In the following, we first introduce the benchmarks we consider to assess the performance of our CtM solution (\Sec{ml}), and the simulator we developed for performance evaluation (\Sec{archi}). Subsequently, we  compare our CtM strategy against the  benchmark called MaxRate, which uses simulated annealing, (\Sec{peva_ctm}),  and then against two ML-based benchmarks (\Sec{peva-ml}). 
The main parameters of our reference scenarios are summarized in \Tab{params}.

\begin{table}
\caption{Simulation parameters
\label{tab:params}
} 
\centering
\begin{tabularx}{.7\columnwidth}{|X|r|}
\multicolumn{2}{>{\hsize=\dimexpr2\hsize+2\tabcolsep+\arrayrulewidth\relax}X}{\centering \bf Indoor factory scenario} \\\hline
Hall size & 80$\times$20$\times$8\,m$^3$\\
\hline
PoA height & 7\,m (3\,GHz), 6\,m (5\,GHz)\\
\hline
User height & 1.5\,m\\
\hline
Clutter height & $2$ m\\
\hline
Clutter density & 40\%\\
\hline
No.\,of PoAs & 8\\
\hline
Frequency & 3\,GHz (PoAs 1--2)\\
& 5\,GHz (PoAs 3--8)\\
\hline
Bandwidth & 20\,MHz\\
\hline
No.\,of end users & 100\\
\hline
No.\,of humans & 200\\
\hline
Required rate & 100\,Mbit/s\\
\hline
$\text{SAR}_\text{wb}$ limit & 80\,mW/kg~\cite{international2020guidelines}\\
\hline
\end{tabularx}

\vspace{3mm}
\begin{tabularx}{.7\columnwidth}{|X|r|}
\multicolumn{2}{>{\hsize=\dimexpr2\hsize+2\tabcolsep+\arrayrulewidth\relax}X}{\centering \bf Urban canyon scenario} \\
\hline
Scenario size & 800$\times$40 m$^2$\\
\hline
PoA height & 10\,m\\
\hline
User height & 0.9\,m or 1.5\,m\\
\hline
No.\,of PoAs & 8\\
\hline
Frequency & 3.5\,GHz (PoAs 1--2)\\
& 5.2\,GHz (PoAs 3--8)\\
\hline
Bandwidth & 20\,MHz\\
\hline
No.\,of end users & 100\\
\hline
No.\,of humans & 200\\
\hline
Required rate & 100\,Mbit/s\\
\hline
$\text{SAR}_\text{wb}$ limit & 80\,mW/kg~\cite{international2020guidelines}\\
\hline
\end{tabularx}
\end{table}

\subsection{MaxRate and ML-based benchmarks}
\label{sec:ml}

We compare the performance of the proposed CtM solution against three different benchmarks. 

The first we consider, called MaxRate,  uses simulated annealing~\cite{schneider2007stochastic} to find a solution that maximizes the minimum among the data rates achieved by devices (\Sec{peva_ctm}), i.e., it addresses the following objective: 
\begin{equation}
\nonumber
\max_{y,P_\text{tx},\phi,\theta,\omega}\min_{d\in\Dc} \mathsf{Rate}(y,P_\text{tx},\phi,\theta,\omega,d).
\end{equation}
By doing so, MaxRate can mimic the traditional approach of improving the overall network performance while guaranteeing fairness among devices.

\begin{figure}
\centering
\includegraphics[width=1\columnwidth]{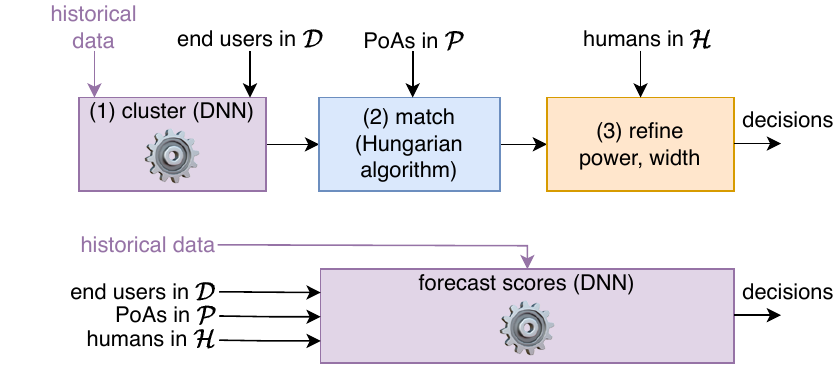}
\caption{
The ML-based benchmarks we compare against: DtM (top), replacing stage~1 of CtM with DNN-based clustering~\cite{ohi2020autoembedder}, and ML-only (bottom), using a DNN to make all decisions. Additions and changes with respect to CtM (\Fig{ctm}) are marked in purple.
    \label{fig:ml}
} 
\end{figure}

Furthermore,  we compare the performance of CtM  against two ML-based benchmarks, whose structure is summarized in \Fig{ml}. 
The first benchmark, depicted in \Fig{ml}(top), is called DNN-then-Match, or DtM for short. It essentially replaces the first stage of CtM, i.e., clustering, with a feed-forward DNN based upon~\cite{ohi2020autoembedder}. The DNN is trained on historical data, i.e., past decisions and their outcome, and returns a user-to-cluster assignment that should result in the best performance.

The second benchmark, presented in \Fig{ml}(bottom), is called ML-only, and reflects the recent trend towards {\em model-free} approaches where decisions are made with  little contribution from domain-specific knowledge or algorithms.

Importantly, both ML-based benchmarks require additional information, i.e., the historical data marked in purple in \Fig{ml}, and must be trained before they can be used. Depending upon the concrete case at hand, historical data may or may not be available, and devoting the necessary resources and time to training may or may not be practical.

\subsection{System simulator}
\label{sec:archi}

\begin{figure}
\centering
\includegraphics[width=.9\columnwidth]{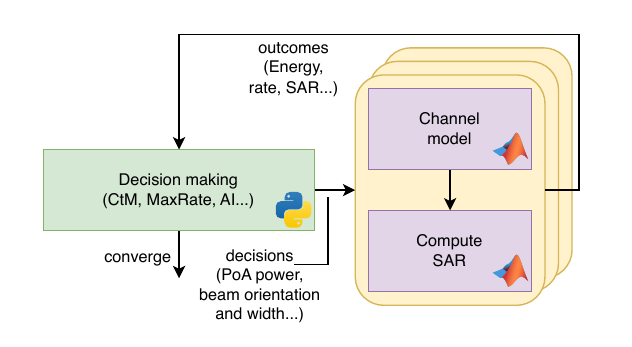}
\caption{
Scheme of the simulator used for performance evaluation.
    \label{fig:archi}
} 
\end{figure}

\begin{figure*}
\centering
\includegraphics[width=.32\textwidth]{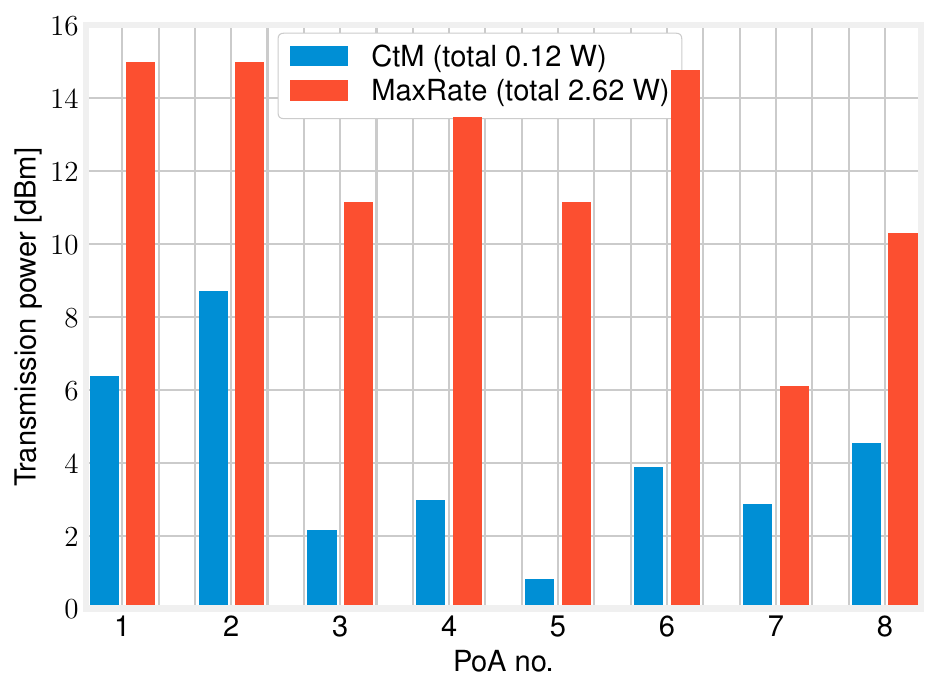}
\includegraphics[width=.32\textwidth]{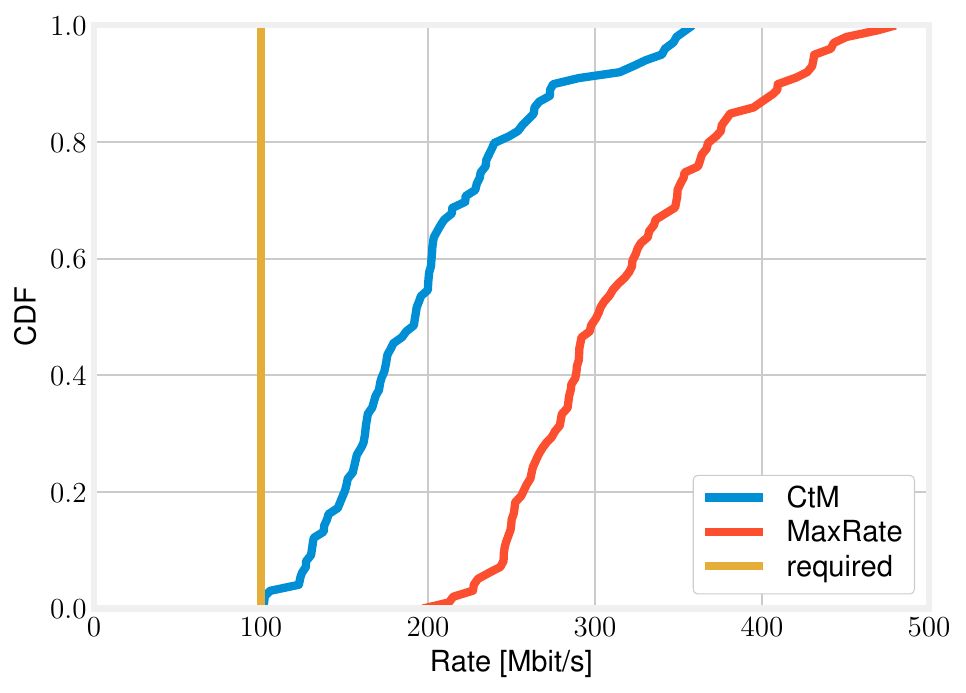}
\includegraphics[width=.32\textwidth]{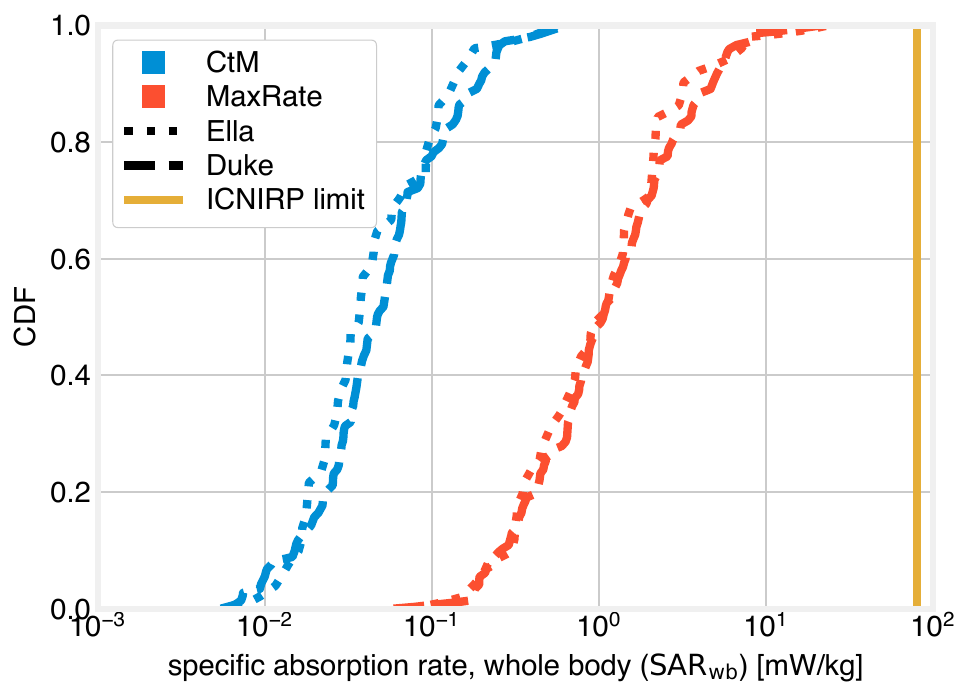}
\caption{
Indoor factory scenario, CtM and the MaxRate benchmark: transmitted power (left), distribution of data rates (center), and $\text{SAR}_\text{wb}$ values (right).
    \label{fig:general}
}
\centering
\includegraphics[width=.45\textwidth]{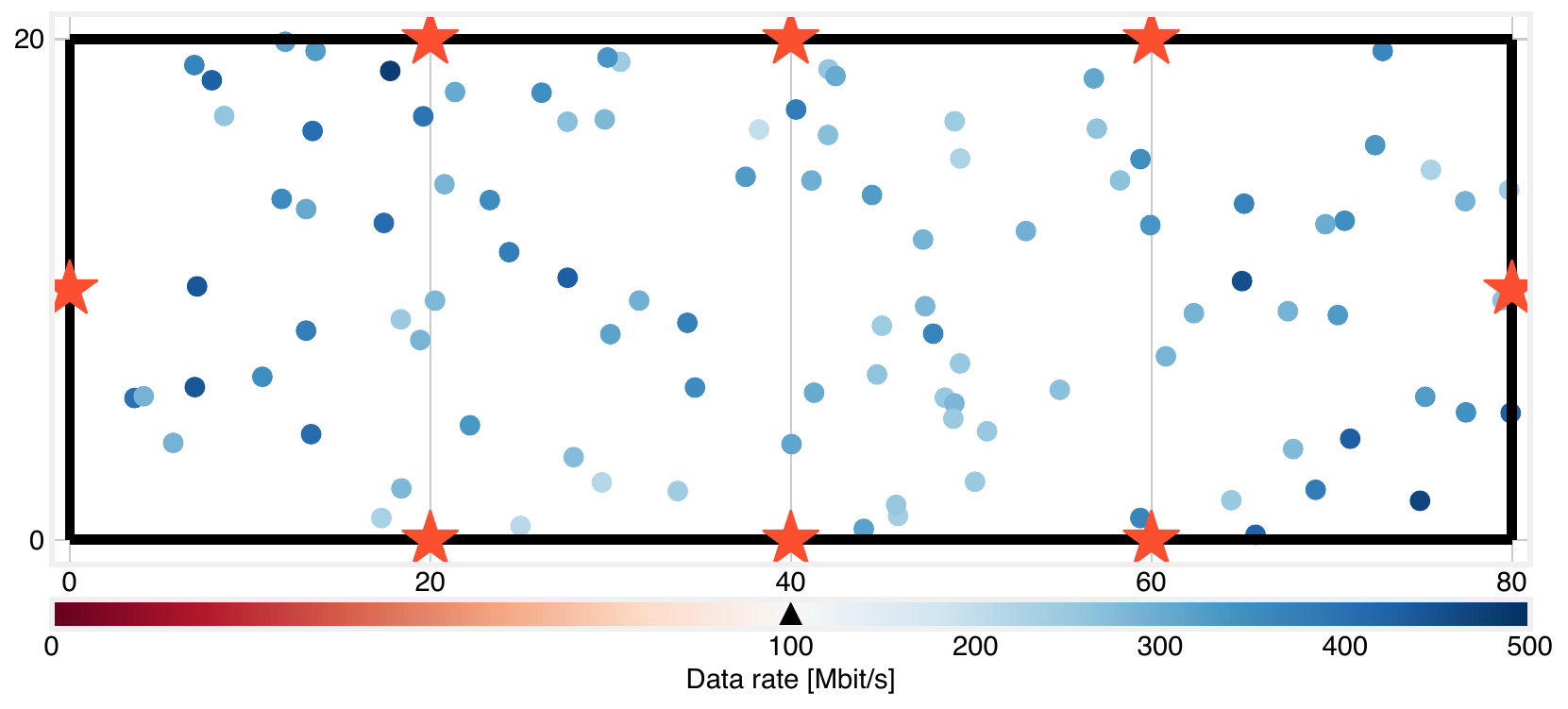}
\includegraphics[width=.45\textwidth]{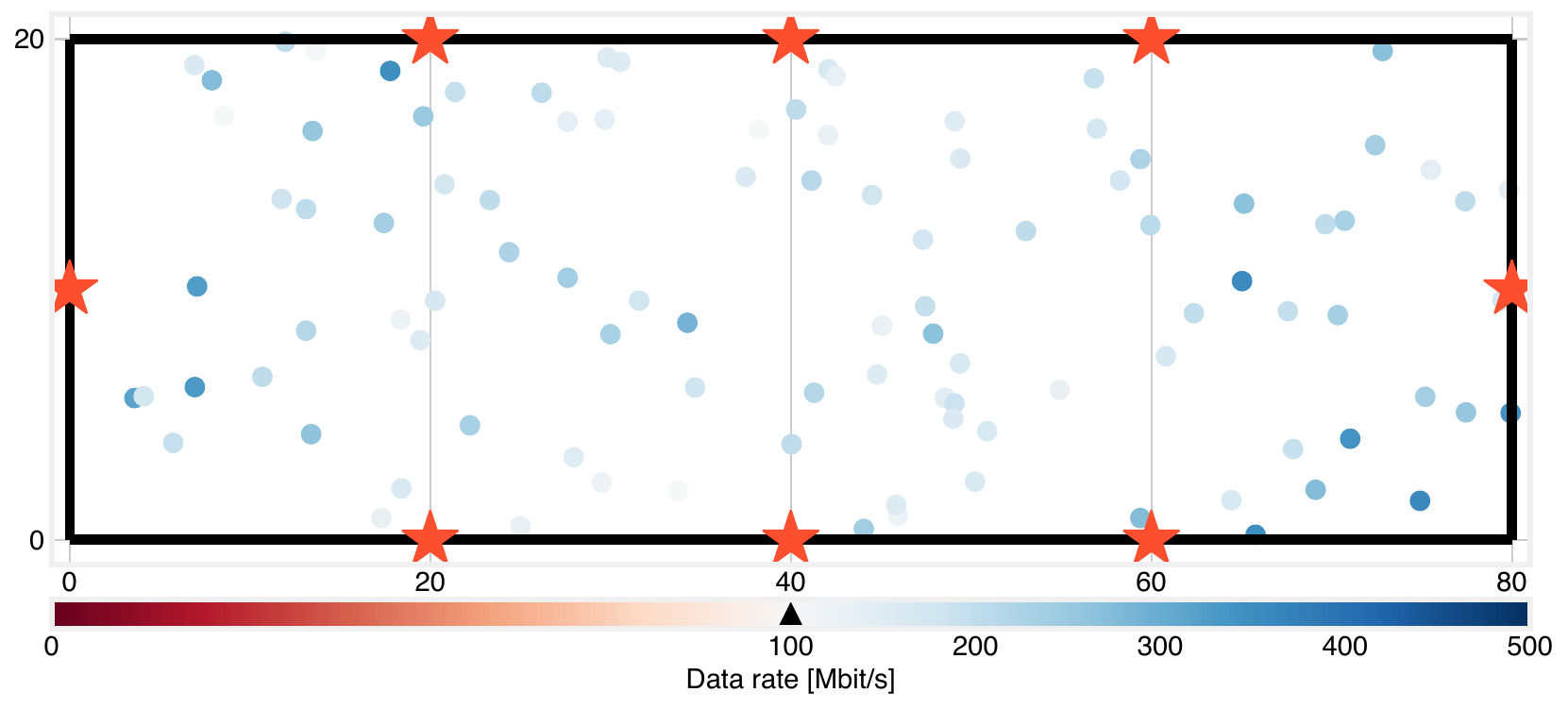}
\caption{
Indoor factory scenario: data rate experienced by different end users under the MaxRate (left) and CtM (right) strategies. The black marker on the color bar corresponds to the required rate. Red stars represent PoAs.
    \label{fig:maps-rate}
}
\centering
\includegraphics[width=.45\textwidth]{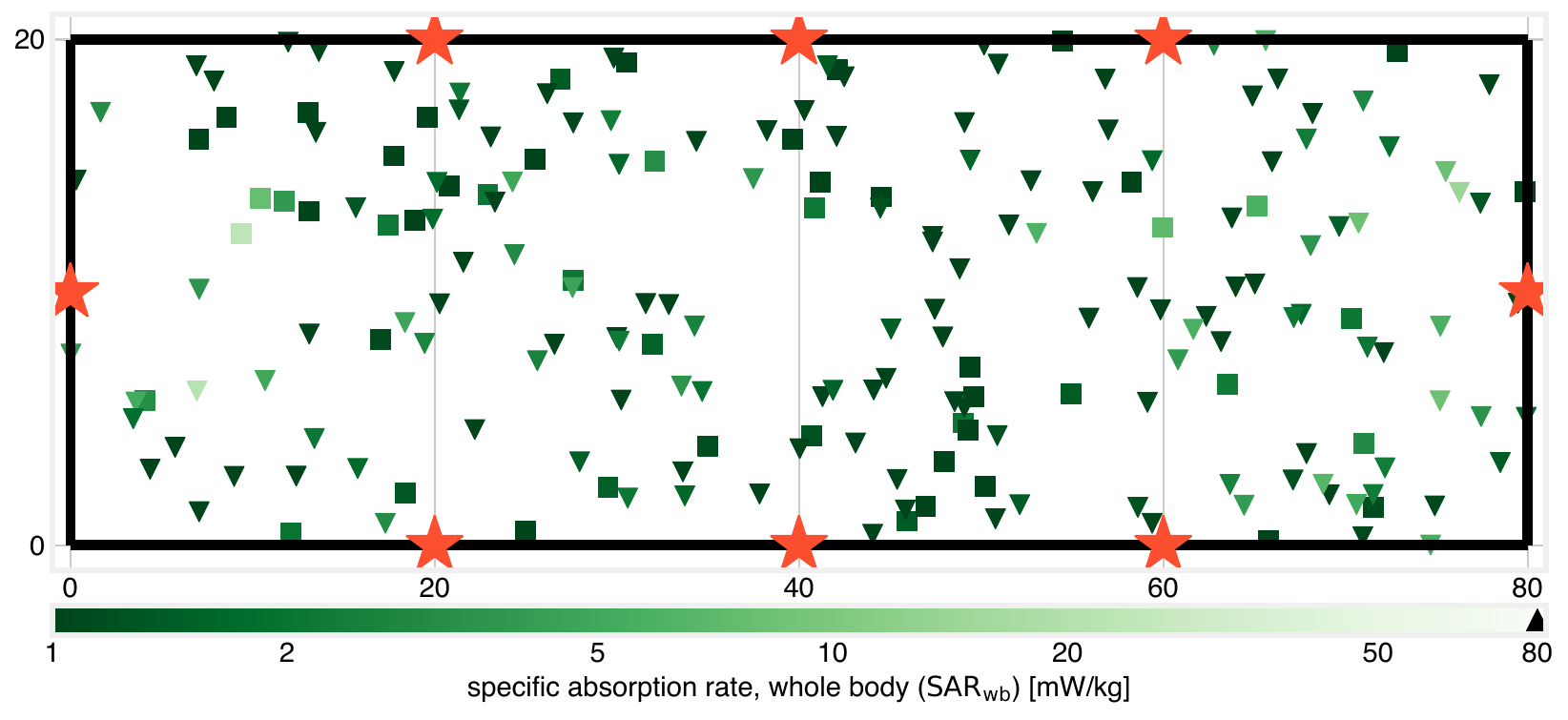}
\includegraphics[width=.45\textwidth]{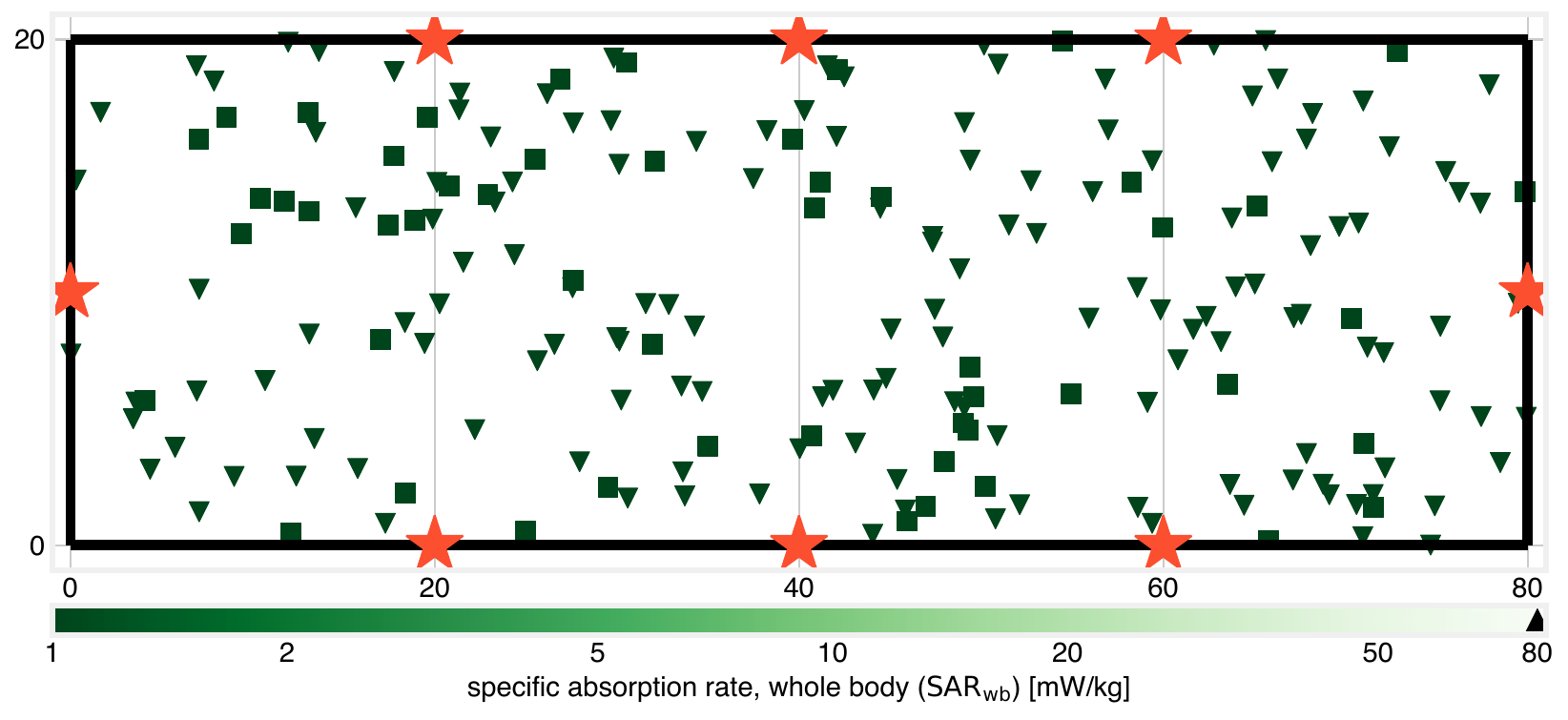}
\caption{
Indoor factory scenario: $\text{SAR}_{\text{wb}}$ experienced by different humans under the MaxRate (left) and CtM (right) strategies. The black marker on the color bar corresponds to the ICNIRP limit. Square and triangle markers correspond (resp.) to humans associated with the Duke and Ella models. Red stars represent PoAs.
    \label{fig:maps-sar}
}
\end{figure*}

For our performance evaluation we leverage an ad-hoc simulator, whose architecture is summarized in \Fig{archi} and includes three main components:
\begin{itemize}
    \item a decision-making module, coded in Python and implementing the CtM and MaxRate strategies;
    \item an implementation of the channel model described in \Sec{scenario}, coded in Matlab;
    \item a module computing EMF exposure values as described in \Sec{exposure}, coded in Matlab.
\end{itemize}

The architecture is highly decoupled: modules run as separate processes, can be -- indeed, are -- coded in different languages, and communicate by exploiting local or remote inter-process communication (IPC) mechanisms.
Such decoupling has the distinct advantage of naturally exploiting the multi-core capabilities of modern hardware; indeed, since each module runs as a separate process, they can efficiently exploit different processors. Even more important, multiple instances of the channel model and SAR computation values can be ran in parallel, so as to evaluate multiple decisions at the same time. This is especially useful for techniques based upon simulated annealing or Markov decision processes, where {\em groups} of alternatives (``population'', ``arms''...) must be evaluated concurrently.

Notably, our decoupled architecture makes it trivial to replace any module with a different implementation thereof; as an example, we might have:
    (a) a decision-making module based on Markov decision processes or reinforcement learning;
    (b)  a different channel model;
   (c) a different way to quantify EMF exposure, using alternative metrics and/or models.
%
On the negative side, the communication and coordination between different processes inevitably results in an overhead. However, compared to the running time of individual modules (especially the ones implementing channel models), such overhead is negligible. It can also be further reduced by using modern IPC techniques like message passing.

We take advantage of our decoupled architecture when implementing the ML-only benchmark, and replace the whole CtM block with a feed-forward DNN taking as an input the decisions (beam orientation, width, and power) and the resulting performance. We train the DNN over a dataset including 10,000~examples of past decisions, and then use it to obtain -- in a generative fashion -- the set of decisions deemed to yield the best performance.

\begin{figure*}
\centering
\begin{minipage}{.32\textwidth}
\centering
\includegraphics[width=1\textwidth]{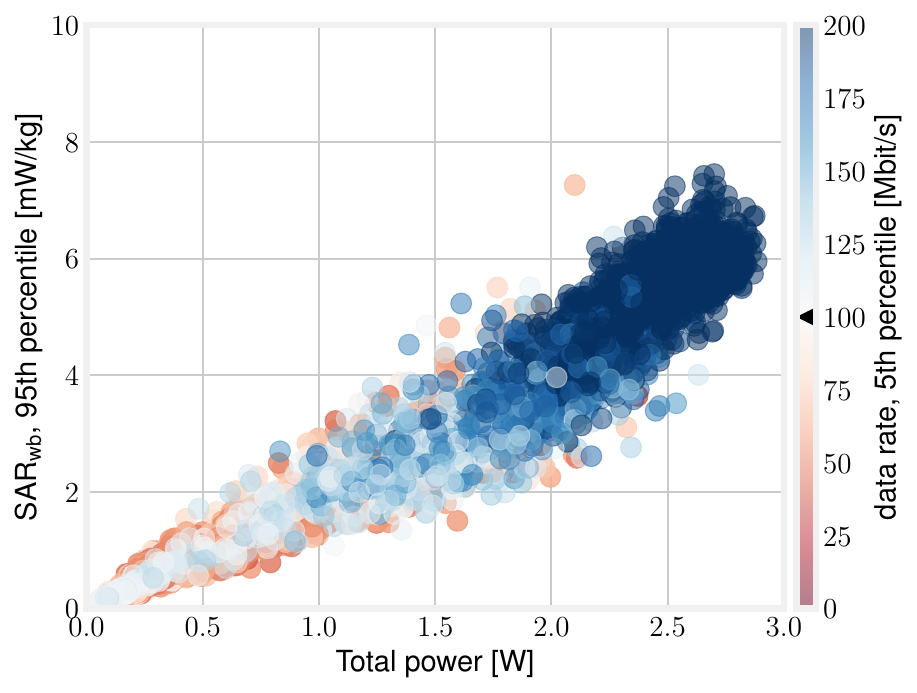}
\caption{
Indoor factory scenario, all solutions considered by all strategies: relationship between total emitted power and 95th percentile of $\text{SAR}_{\text{wb}}$, with the color of each marker corresponding to the rate experienced by the 5th percentile of end users.
    \label{fig:tradeoffs}
}
\end{minipage}
\hspace{.2em}
\begin{minipage}{.64\textwidth}
\centering
\includegraphics[width=.49\textwidth]{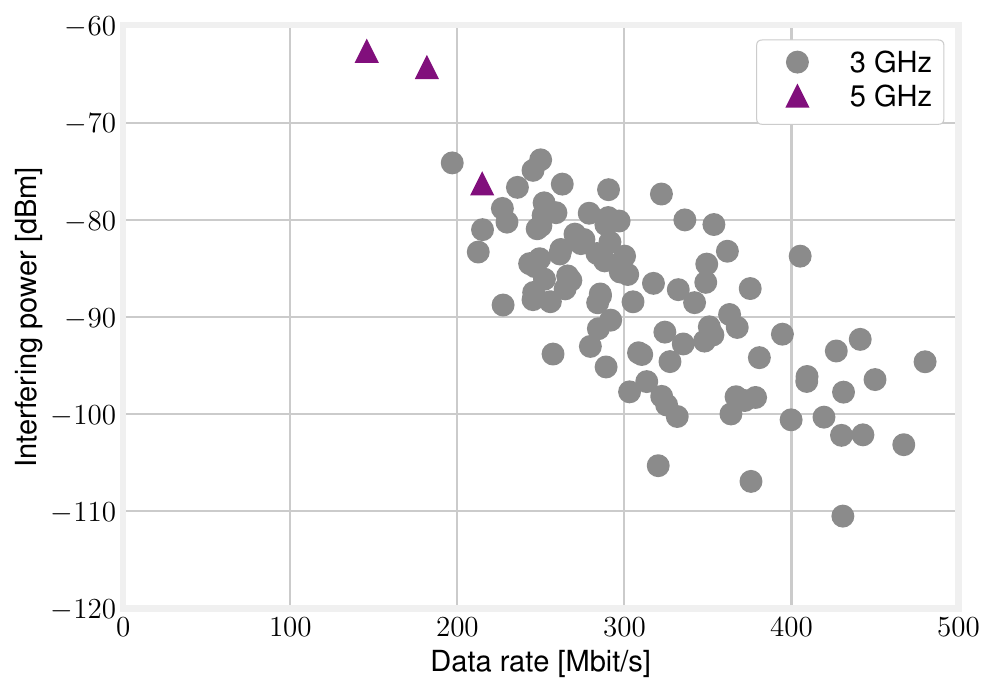}
\includegraphics[width=.49\textwidth]{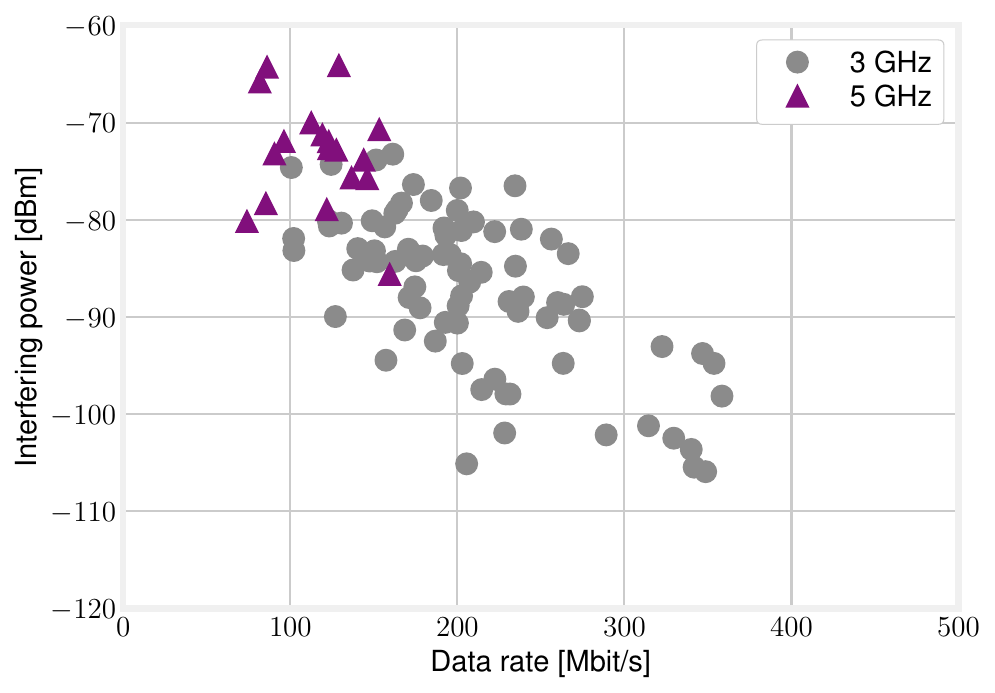}
\caption{
Indoor factory scenario: data rate (x-axis) and interfering power (y-axis) experienced by all end users under the MaxRate (left) and CtM (right) strategies. The color of each marker denotes the frequency of the PoA serving the corresponding end user.
    \label{fig:freqs}
}
\end{minipage}
\end{figure*}

\begin{figure*}
\centering
\includegraphics[width=.32\textwidth]{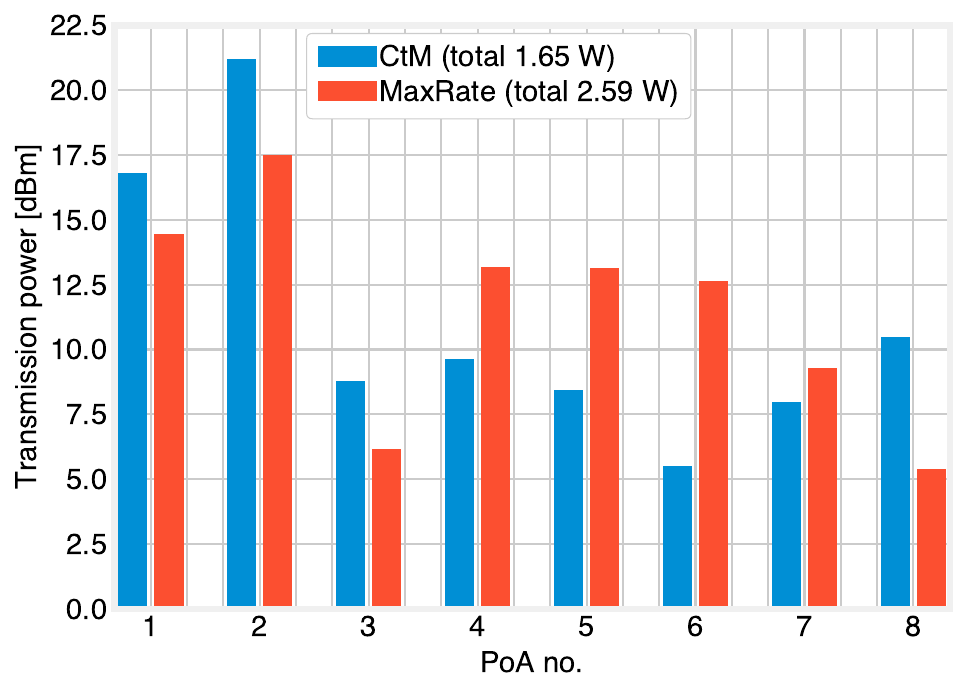}
\includegraphics[width=.32\textwidth]{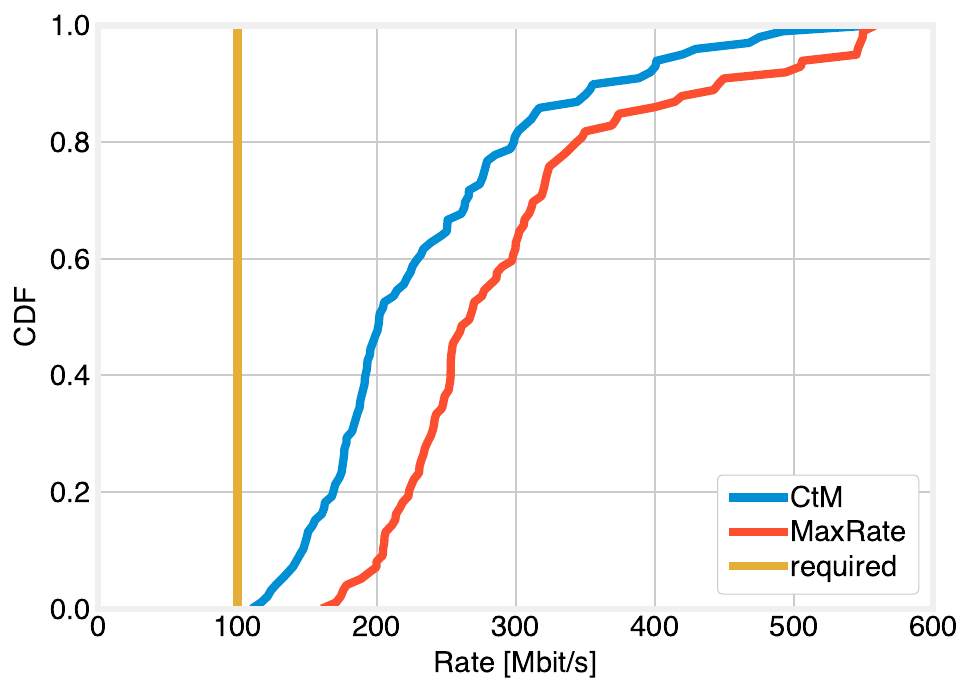}
\includegraphics[width=.32\textwidth]{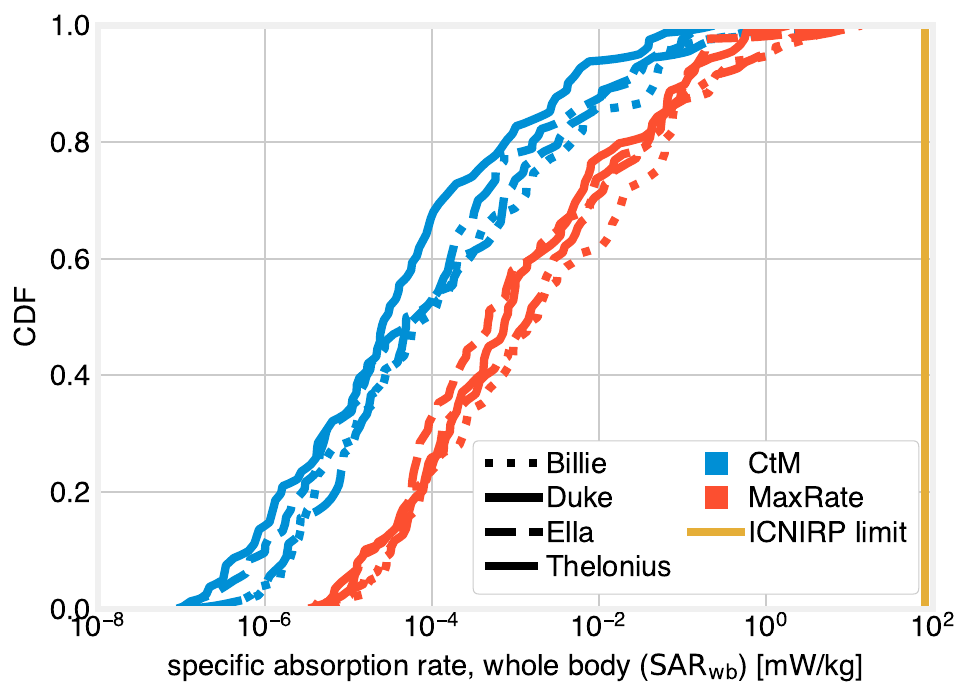}
\caption{
Urban canyon scenario, CtM and the MaxRate benchmark: transmitted power (left), distribution of data rates (center), and $\text{SAR}_\text{wb}$ values (right).
    \label{fig:general2}
}
\centering
\includegraphics[width=.45\textwidth]{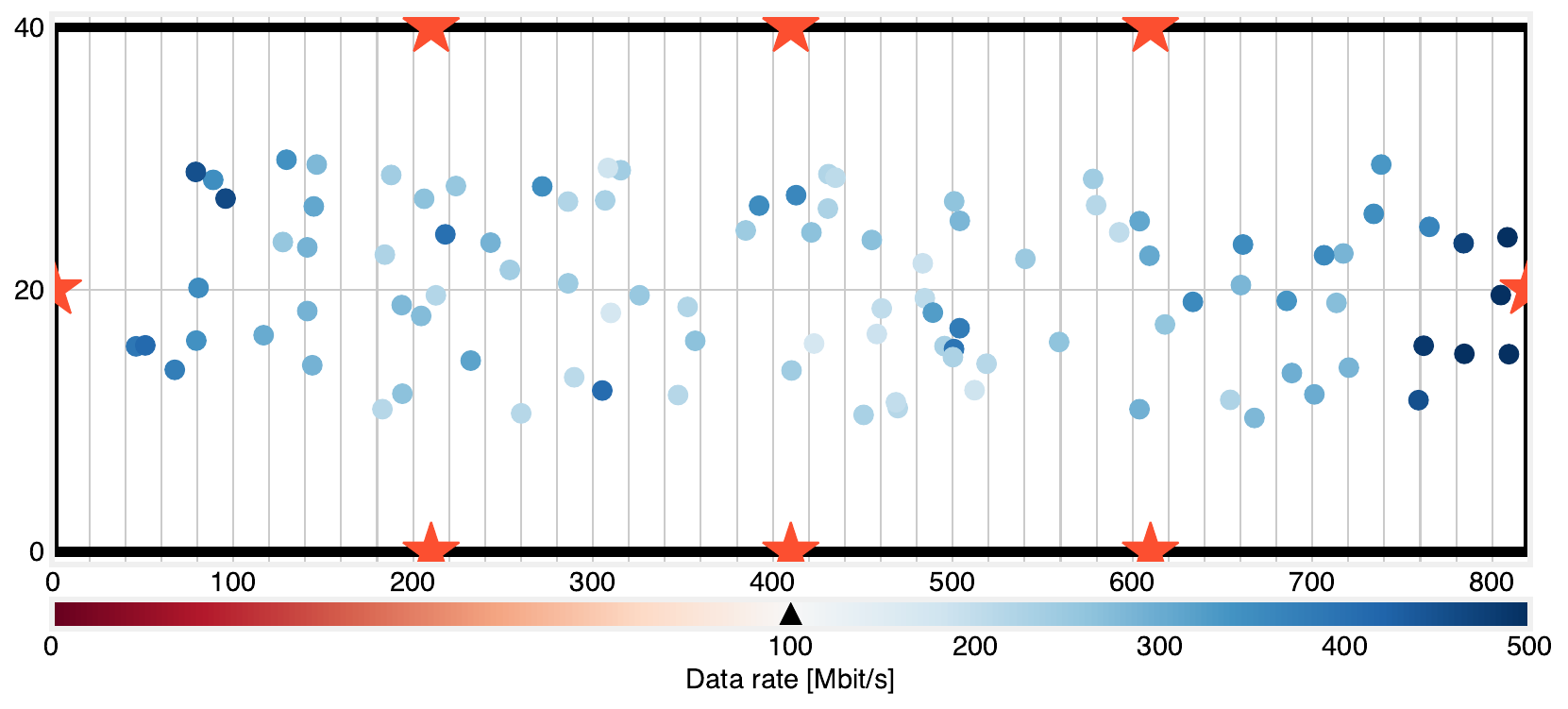}
\includegraphics[width=.45\textwidth]{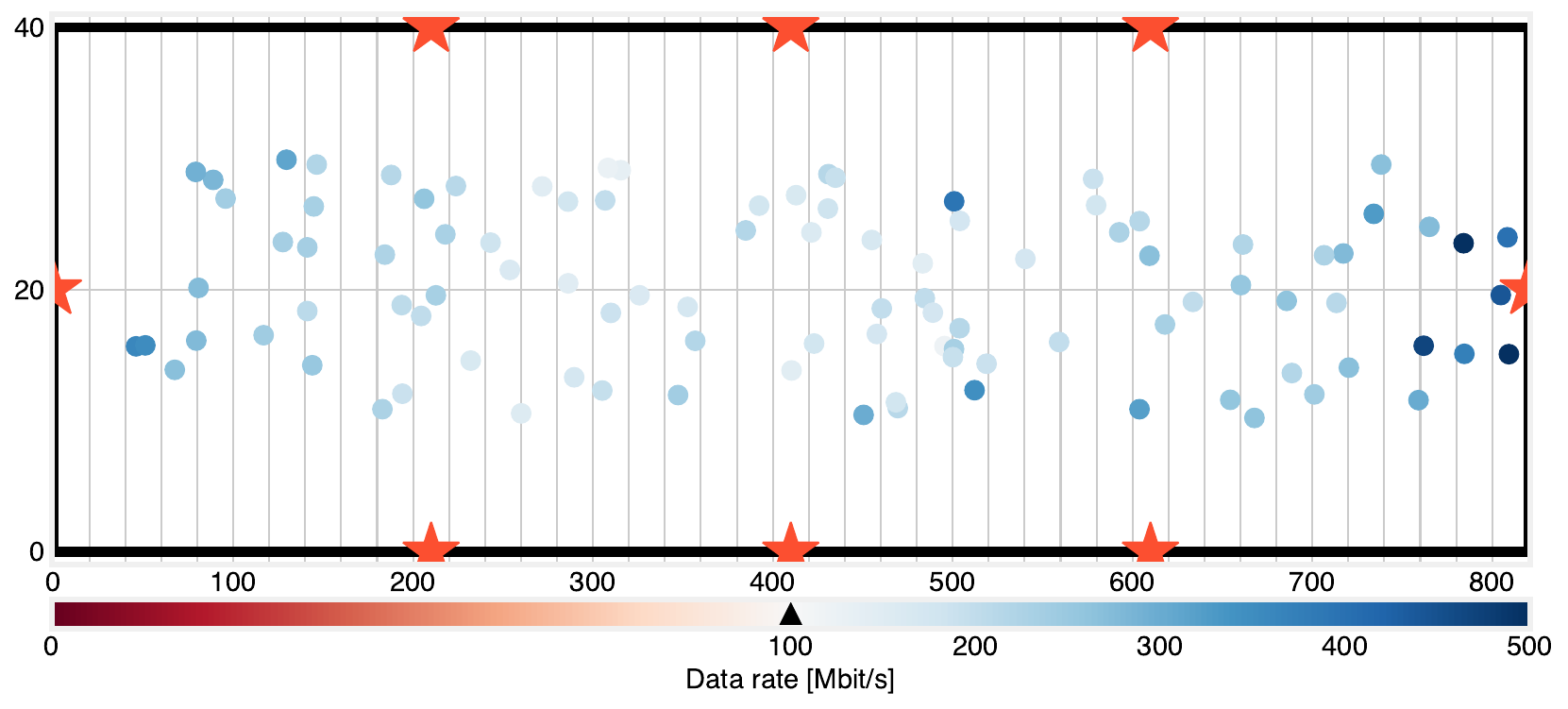}
\caption{
Urban canyon scenario: data rate experienced by different end users under the MaxRate (left) and CtM (right) strategies. The black marker on the color bar corresponds to the required rate. Red stars represent PoAs.
    \label{fig:maps-rate2}
}
\centering
\includegraphics[width=.45\textwidth]{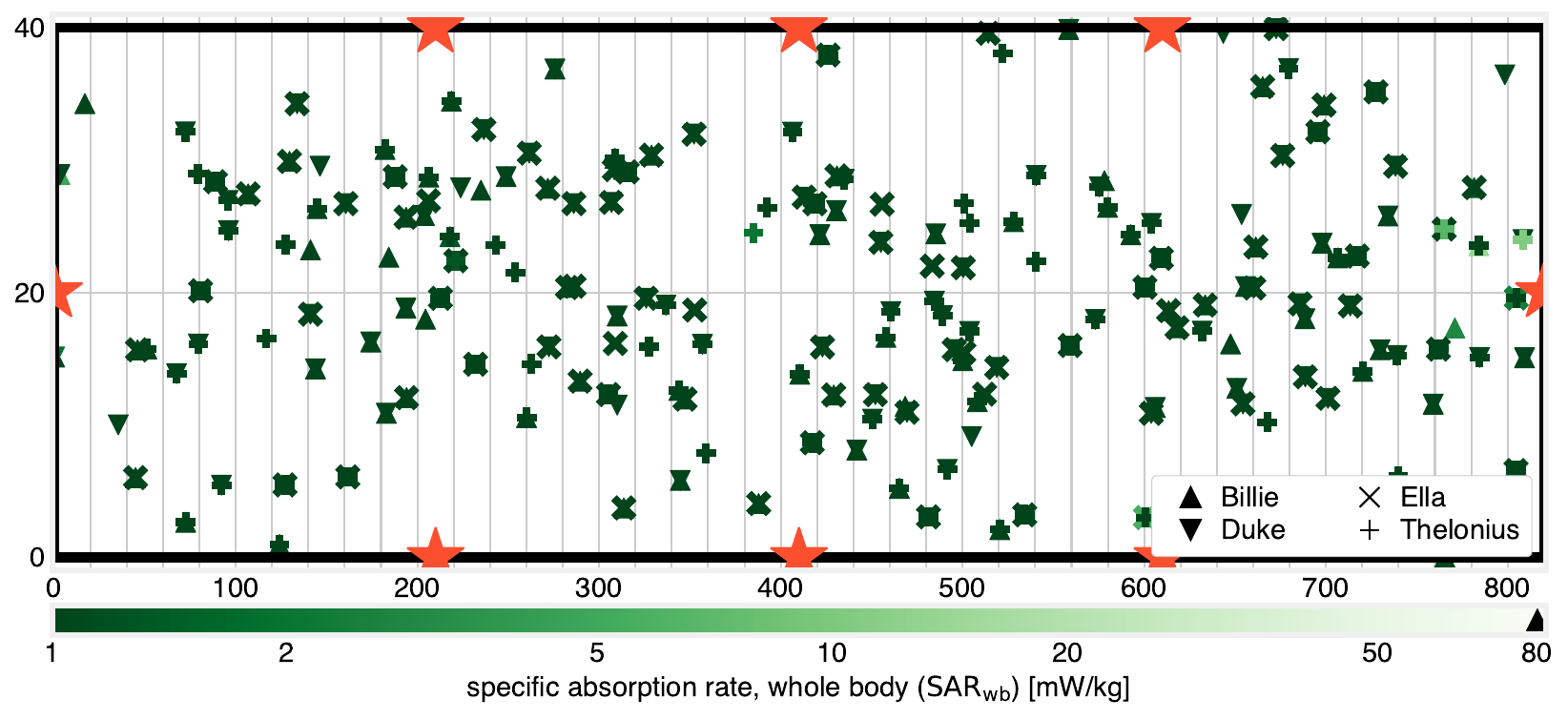}
\includegraphics[width=.45\textwidth]{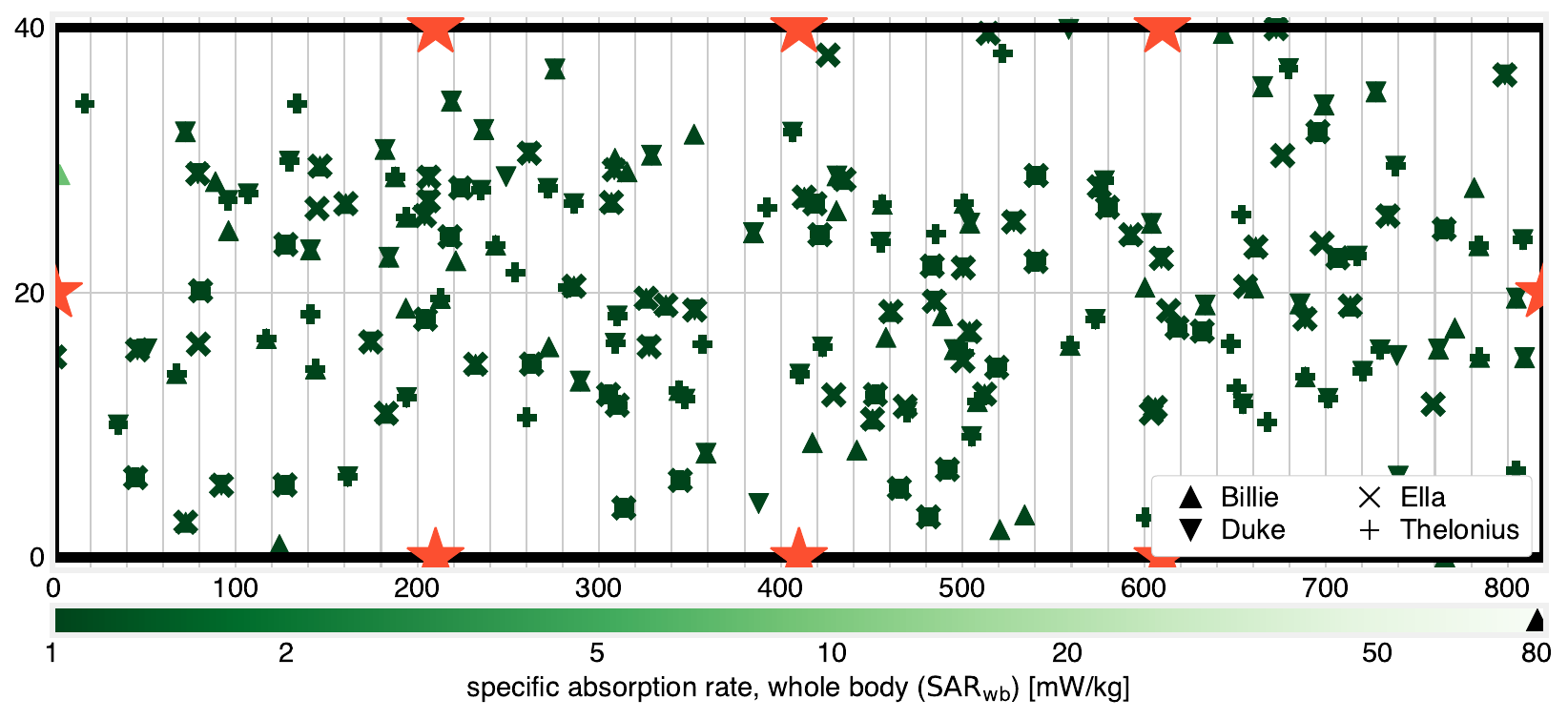}
\caption{
Urban canyon scenario: $\text{SAR}_{\text{wb}}$ experienced by different humans under the MaxRate (left) and CtM (right) strategies. The black marker on the color bar corresponds to the ICNIRP limit. Square and triangle markers correspond (resp.) to humans associated with the Duke and Ella models. Red stars represent PoAs.
    \label{fig:maps-sar2}
}
\end{figure*}

\subsection{CtM performance against MaxRate}
\label{sec:peva_ctm}

We now 
study the performance of the basic CtM algorithm described in \Sec{algo} and compare it to the MaxRate benchmark in the two scenarios described in \Sec{scenarios}.

\noindent{\bf Indoor factory scenario.}
We begin by investigating the most basic aspect of CtM's and MaxRate's performance, i.e., the incurred energy consumption. Concerning this critical aspect, \Fig{general}(left) shows that CtM outperforms MaxRate, yielding a power consumption that is almost an order of magnitude lower. Groups of bars in the plot correspond to different PoAs, and confirm that CtM can significantly reduce the transmission power of all PoAs compared to MaxRate. Also notice how, for both strategies, PoAs~1 and~2, operating at 3~GHz and typically serving  farther-away end users, get assigned higher values of transmission power than the other PoAs that  operate at 5~GHz.

Striking as \Fig{general}(left) is, one might rightfully wonder whether such a reduced energy consumption comes at a cost in terms of data rate and service quality. The answer, as shown in \Fig{general}(center), is both positive and negative. Indeed, under the CtM strategy (blue line in the plot) end users get lower data rates than under MaxRate (red line in the plot); however, such data rates are always (and sometimes significantly) above the required level (yellow line in the plot). This is consistent with the way data rates are included in our problem formulation in \Sec{problem} and, more specifically, with constraint \Eq{rat}: so long as all end users are guaranteed~$\textsf{Rate}^{\min}(d)$, there is no reason to further increase data rates. At a more general level, \Fig{general}(left) and \Fig{general}(center) suggest how adopting a human-centric approach, hence, using data rate performance as a constraint and not an objective, can bring substantial energy savings without jeopardizing service requirements.

In \Fig{general}(right), we move to the other human-centric metric we consider, i.e., EMF exposure as quantified through $\text{SAR}_{\text{wb}}$. Once can immediately see that CtM yields a much lower $\text{SAR}_{\text{wb}}$ than MaxRate; further, $\text{SAR}_{\text{wb}}$ levels for either strategies are significantly below the limit values recommended in~\cite{international2020guidelines}, to wit, 80~mW/kg.
It is perhaps even more interesting to remark how there are {\em two} lines in the plot for each strategy, one dashed and one dotted, corresponding (resp.) to the Ella and Duke models introduced in \Sec{exposure}. This further confirms that our approach can account for individual characteristics when assessing~$\text{SAR}_\text{wb}$.

We now check whether there is any clear space pattern in the distribution of rate by plotting, in \Fig{maps-rate}, the location of each end user  and the rate they get under the MaxRate and CtM strategies. Consistently with \Fig{general}(center), MaxRate ensures to virtually all end users a data rate that is much higher than required, hence, all markers in the left-hand side map are deep blue. In the right map, instead, we can see many lighter markers, though no red ones -- highlighting the fact that CtM yields rate values that are closer to, but above, the required one. As one might expect, nodes with lower rate tend to be farther away from the PoA serving them, hence, experience higher attenuation and/or interference.

Similarly, \Fig{maps-sar} presents the $\text{SAR}_{\text{wb}}$ levels experienced by humans in~$\Hc$, under the MaxRate and CtM solutions. As we can expect from \Fig{general}(right), CtM results in uniformly low $\text{SAR}_{\text{wb}}$ values, hence, dark green markers. MaxRate, on the other hand, results in higher $\text{SAR}_{\text{wb}}$ levels, hence, slightly lighter markers. As per \Fig{general}(right), however, all $\text{SAR}_{\text{wb}}$ values are significantly lower than the limit, thus, there are no purple markers on the plot. Also notice how the shape of the marker identifies the model (Ella or Duke) used to determine the $\text{SAR}_{\text{wb}}$ of that specific human.

In \Fig{tradeoffs}, we look a little deeper into the relationship between the transmission power selected by PoAs and the $\text{SAR}_{\text{wb}}$ level experienced by humans. Each marker in the plot corresponds to a solution (including infeasible ones) considered by any of the strategies. Its positions along the $x$- and $y$-axes correspond, respectively, to the total power level and 95th percentile of $\text{SAR}_{\text{wb}}$ (computed over all humans). We can observe a very strong correlation between these two quantities.
Even more interestingly, from the color of the markers, one can see that solutions with high power are almost invariably feasible. On the other hand, low power levels are associated with  many infeasible solutions but also -- critically -- some feasible ones. Indeed, CtM can explore that part of the solution space, and find solutions that are consistent with the data requirements while incurring little energy consumption and EMF exposure.

Finally, \Fig{freqs} sheds additional light on how PoAs operating at different frequencies are employed by the MaxRate and CtM strategies. Each marker in the plots corresponds to an end user; its position along the $x$- and $y$-axes corresponds, respectively, to the achieved data rate and the incurred interfering power; marker colors represent  the frequency of the PoA serving the associated end user. We can observe that the higher frequency (5~GHz, purple markers) is associated with higher interfering power and lower data rates. Nonetheless, CtM (right plot) can leverage such a frequency to serve a much larger number of end users, while MaxRate (left plot) leverages almost exclusively the lower, 3~GHz frequency (grey markers in the plots). As also highlighted in \Fig{general}(left), however, intensively using that frequency necessitates higher transmission power, which results in higher energy consumption and EMF exposure.

\noindent{\bf Urban canyon scenario.}
We now move to the urban canyon scenario, and assess whether the behavior of CtM and MaxRate follows the same behavior as in the indoor factory.

\Fig{general2}(left) confirms the behavior we observed in \Fig{general}(left), with CtM resulting in a significantly lower power consumption than MaxRate. However, we can notice that the power used by CtM is higher than in the indoor factory scenario (\Fig{general}(left)); the main reason for this difference is the larger size of the urban canyon scenario, which results in longer distances between PoAs and users, hence, higher attenuation. By looking at the columns reporting the power used by individual PoAs, we can also observe how CtM tends to use PoAs~1 and~2, placed on the two ends of the road stretch, to a larger extent than MaxRate. This is due to the elongated shape of the urban canyon scenario: serving many users through PoAs 3--8 (placed along the long sides of the area under study) would require using wider beams, which would result in more energy (and EMF exposure); on the other hand, PoAs~1 and~2 (placed on the short sides of the area) can serve many users with narrower beams.

As one might expect, CtM's lower power and reliance on the lower-frequency PoAs result in a significantly lower rate than MaxRate, as shown in \Fig{general2}(center). However, all rate values are consistent with the 100~Mbit/s~requirement (denoted by a yellow line in the plot), hence, CtM's solutions are feasible. Importantly, this is essentially the same behavior as in \Fig{general}(center); in both scenarios, CtM can provide users with the required data rate while minimizing power consumption, configuring the PoAs in the way that best suits the scenario at hand.

Consistently with \Fig{general2}(center), and similarly to \Fig{general}(right), \Fig{general2}(right) shows that CtM yields a much lower exposure than MaxRate. This is true for all phantoms (notice that all four are used for the urban canyon scenario), even though individual phantoms do have different exposure distributions. Finally, similarly to \Fig{general}(right), the EMF exposure is always significantly smaller than the ICNIRP reference level, marked with a yellow line in the plot.

\Fig{maps-rate2}, depicting the data rate experienced by each network user, further highlights the different way in which MaxRate (left map) and CtM (right map) use the PoAs. The difference is especially clear by considering the users around the middle of the topology: as CtM mostly uses PoAs 1--2, those users consistently have a lower rate. MaxRate is instead able to give some (not many) users therein higher rates by using the other PoAs, at the cost of a higher energy consumption.

Last, \Fig{maps-sar2} confirms that, consistently with \Fig{general2}(right) and similarly to \Fig{maps-sar}, SAR$_{\rm wb}$ levels experienced by all humans in the topology are comfortably within the ICNIRP reference values. This is true for all phantoms (represented with different markers in the maps), and even for humans in the coverage areas of multiple PoAs, e.g., around the top-left part of \Fig{maps-sar2}(left).

\subsection{CtM performance against ML-based benchmarks}
\label{sec:peva-ml}

Back to the indoor factory scenario, we now compare the performance of CtM against the ML-based benchmarks presented in \Sec{ml}: DtM, which replaces the clustering stage in CtM with a DNN-based block, and ML-only, seeking to replace the whole CtM with a DNN.

\begin{figure}
\centering
\includegraphics[width=.8\columnwidth]{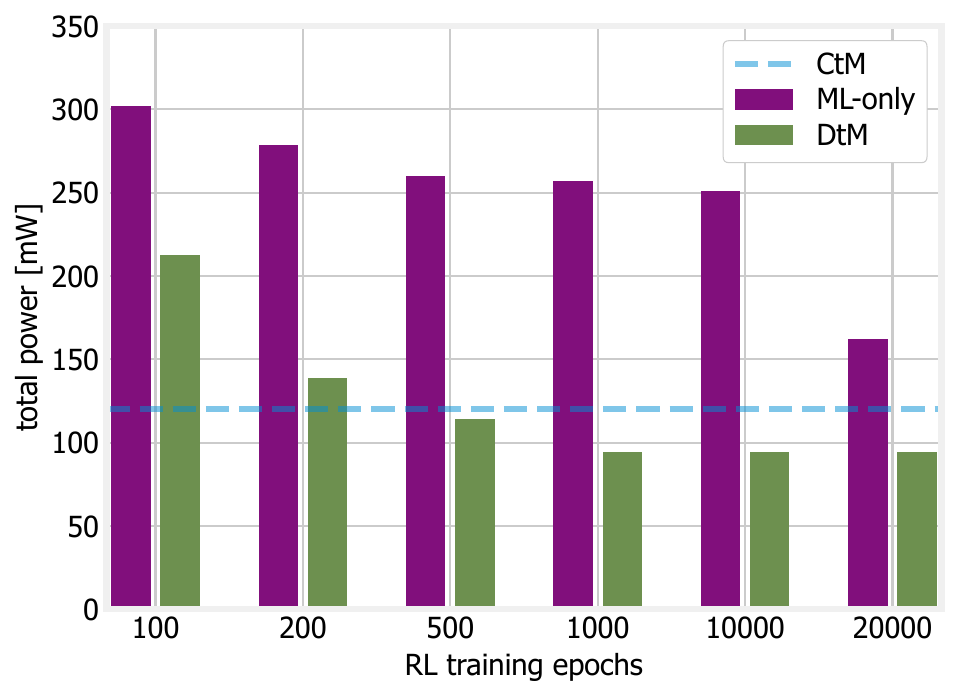}
\caption{
Total power consumption under CtM, DtM and ML-only approaches as a function of the number of epochs allowed for DNN.
    \label{fig:ml-train}
}
\end{figure}

\begin{figure*}
\centering
\includegraphics[width=.32\textwidth]{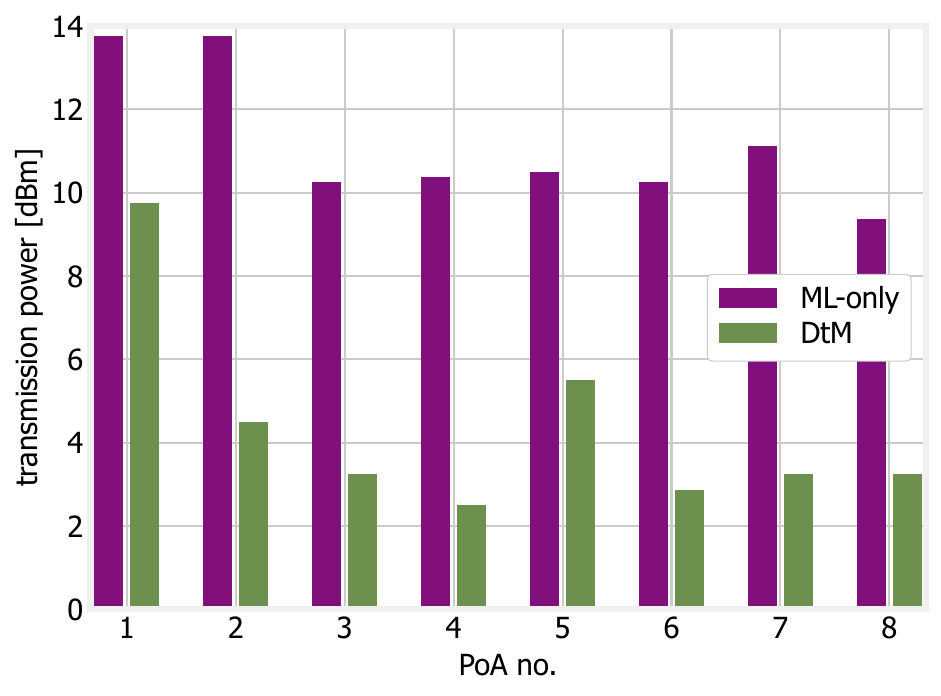}
\includegraphics[width=.32\textwidth]{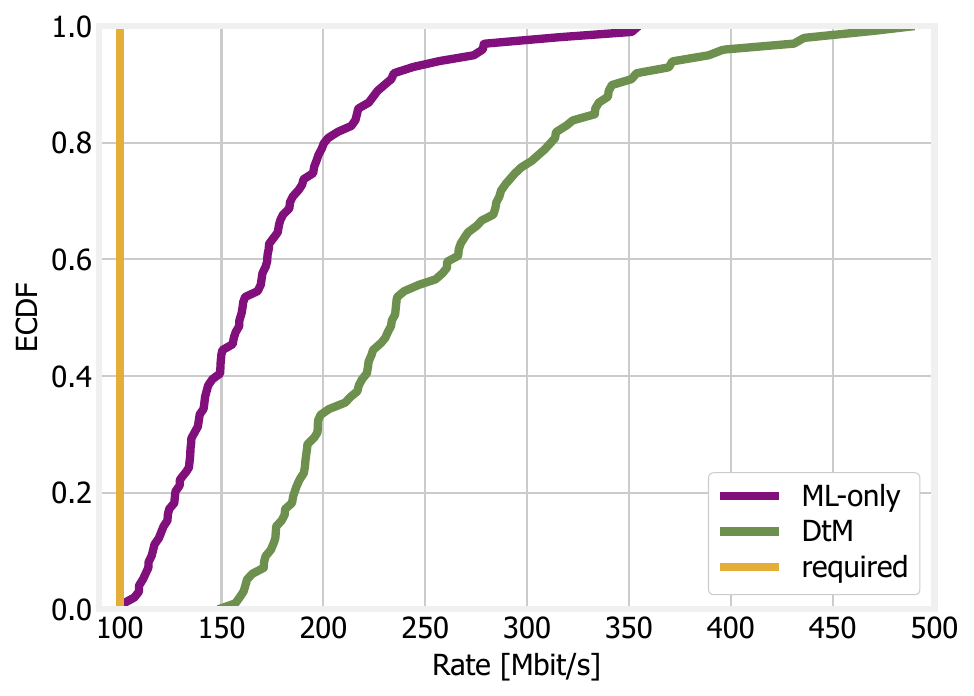}
\includegraphics[width=.32\textwidth]{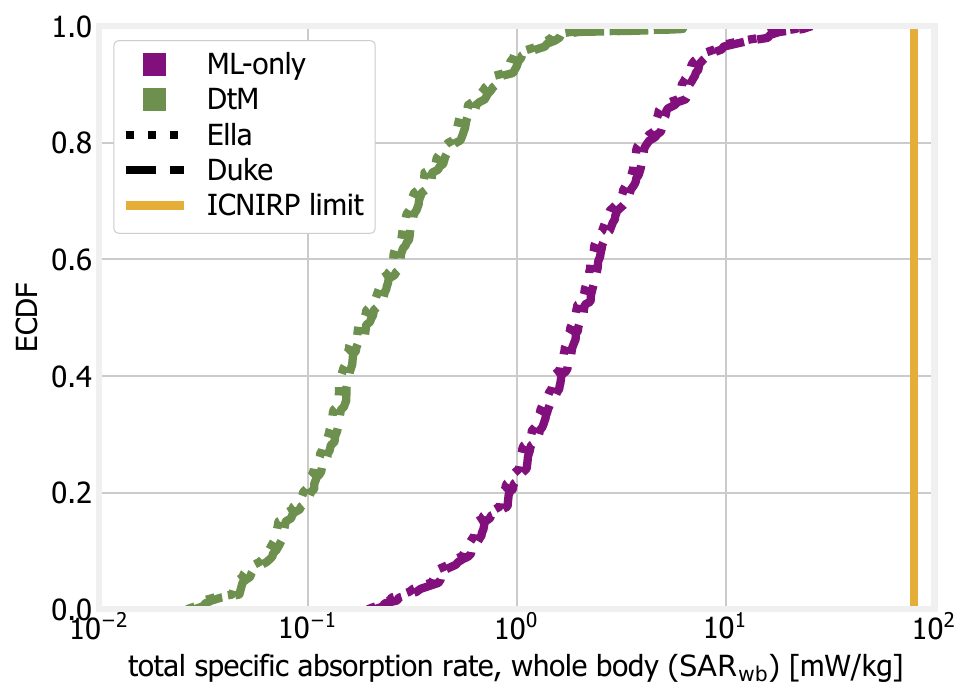}
\caption{
DtM and ML-only for 1,000 training epochs: transmitted power (left), distribution of data rates (center), and $\text{SAR}_\text{wb}$ values (right).
    \label{fig:ml-general}
}
\end{figure*}

\Fig{ml-train} presents the total power consumption yielded by DtM (green bars) and ML-only (purple bars), with the power consumption of the original CtM reported as a blue, dashed line as a reference. We can immediately see that the green bars are consistently lower than the alternatives; indeed, DtM outperforms the original CtM provided that a sufficient number of epochs is allowed for training. As for ML-only, it results in a power consumption that is much higher than the alternatives. The situation improves as more training epochs are performed; however, even after tens of thousands of epochs, ML-only is unable to match CtM.

The results in \Fig{ml-train} confirm the intuition behind the design of DtM: ML is best used to {\em complement} domain-specific knowledge and approaches, as opposed to replacing them. One may conjecture that additional training (i.e., going further to the right in \Fig{ml-train}) might lead to ML-only eventually outperforming CtM; however, the resources and time needed for such training would render that approach infeasible or impractical in many real-world cases.

Last, let us focus on the two ML-based benchmarks, and seek to understand the effect of combining ML with a domain-specific approach, as opposed to replacing the latter with the former. Specifically, \Fig{ml-general}(left) shows that DtM uses substantially lower power levels than ML-only, for all PoAs. This further highlights the effectiveness of retaining the blocks of CtM in charge of making power and width decisions, as per \Fig{ml}(top). For the same reasons, DtM yields a lower exposure than ML-only (\Fig{ml-general}(right)) while  honoring the minimum-rate constraint (\Fig{ml-general}(center)).

\section{Related Work}
\label{sec:relwork}

The  {\em human-centric} expression is applied to 5GB/6G networks in multiple contexts and with different meanings. Networks that are expected to closely interact with humans, e.g., in the tactile Internet, are called human-centric in~\cite{maier20216g}, which investigates the additional capabilities and performance (most notably, latency) required by these applications. \cite{giordani2020toward}~takes a more technical view on the same issue, identifying novel 6G use cases (e.g., telepresence and augmented reality) and the associated requirements. At the same time, the societal impact of 6G networks is evaluated and assessed in~\cite{ystgaard2021bring}, focusing on aspects like empowerment. Finally, since artificial intelligence (AI) is a key part of 6G, research efforts on human-centric networking are looking into making AI explainable~\cite{wang2021applications} and to better control the data used for its training~\cite{saad2019vision}.

A second research area our work is related to is the so-called green networking, exploring the trade-offs between network performance and the energy consumption (hence, carbon footprint) they incur. Most approaches act on the network infrastructure and optimize its configuration, e.g., enlarging the coverage area of some base stations so that others can be switched off~\cite{verma2020toward}. To achieve the former objective, novel transmission techniques are often employed, e.g., passive intelligent reflective surfaces~\cite{yu2021irs}. Many recent approaches leverage AI to either predict the demand for coverage and capacity~\cite{vallero2020processing}, evaluate the impact of switching decisions~\cite{mao2021ai}, or predict  spectrum behavior~\cite{yang2021spectrum}. At the same time, assessing and ensuring the sustainability of AI itself is a highly active research area~\cite{schwartz2020green}, with approaches targeting both deep neural network (DNN) architectures~\cite{du2020green} and distributed learning paradigms~\cite{xue2021eosdnn}.

Finally, several works  try to combine EMF exposure assessment and wireless (mostly, cellular) networking. Many early approaches~\cite{chiaraviglio2021cellular} focus on  network {\em planning}, working under the assumption that the infrastructure is  given and  impossible to control beyond setting power levels. Under such conditions, they check whether EMF limits are likely to be exceeded. More recent works look at network management, seeking to adapt the configuration of  PoAs so as to account for both EMF limits and network performance; as an example, \cite{liesegang2023emf}~focuses on power control in cell-free massive multiple-input-multiple-output  systems.
We seek to further advance with respect to these works by (i) fully modeling and exploiting the configurability of modern-day networks (e.g., beamforming and PoA selection), and (ii) estimating the levels of EMF exposure considering both the presence of users and non-users and their anatomical variability.

Last, a preliminary version of this work has been presented in~\cite{noi-wowmom24}. With respect to our conference paper, this work compares CtM against several benchmarks, including ML-based solutions. Additionally, it  evaluates CtM and its alternatives in a urban canyon network and propagation scenario,  considering new phantoms for EMF exposure assessment.

\section{Conclusions}
\label{sec:future}

We have addressed the scenario of cell-less networks, where there is no fixed, {\em a priori} association between users and PoAs, and management decisions must be made in a human-centric fashion, accounting for network performance as well as power consumption and EMF exposure. In order to cope with the problem complexity, we proposed a solution concept called CtM, for cluster-than-match. As its name suggests, CtM performs two main steps: first clustering end users, and then associating PoAs with clusters. CtM is able to minimize energy consumption; at the same time, CtM meets the rate requirements of all users and honors EMF exposure limits for all humans on the topology, including those who are not network users. Our performance evaluation, which leverages detailed models for EMF exposure estimation and standard-specified signal propagation models, compares CtM's performance against that of traditional and ML-based benchmarks, showing that CtM can reduce the power consumption by over 80\%.

\bibliographystyle{IEEEtran}
\bibliography{refs}

\end{document}